\documentclass[aps,twocolumn,superscriptaddress]{revtex4}
\usepackage{graphicx}
\pdfoutput=1

\newcommand{\bra}[1] {\left\langle #1 \right|}
\newcommand{\ket}[1] {\left| #1 \right\rangle}

\begin{document}

\title{Speed limits for two-qubit gates with weakly anharmonic qubits}

\author{Sahel Ashhab}
\affiliation{Advanced ICT Research Institute, National Institute of Information and Communications Technology (NICT), 4-2-1, Nukui-Kitamachi, Koganei, Tokyo 184-8795, Japan}

\author{Fumiki Yoshihara}
\affiliation{Advanced ICT Research Institute, National Institute of Information and Communications Technology (NICT), 4-2-1, Nukui-Kitamachi, Koganei, Tokyo 184-8795, Japan}

\author{Tomoko Fuse}
\affiliation{Advanced ICT Research Institute, National Institute of Information and Communications Technology (NICT), 4-2-1, Nukui-Kitamachi, Koganei, Tokyo 184-8795, Japan}

\author{Naoki Yamamoto}
\affiliation{Quantum Computing Center, Keio University, 3-14-1 Hiyoshi, Kohoku-ku, Yokohama, Kanagawa 223-8522, Japan}
\affiliation{Department of Applied Physics and Physico-Informatics, Keio University, Hiyoshi 3-14-1, Kohoku-ku, Yokohama 223-8522, Japan}

\author{Adrian Lupascu}
\affiliation{Institute for Quantum Computing, University of Waterloo, Waterloo, ON, Canada N2L 3G1}
\affiliation{Department of Physics and Astronomy, University of Waterloo, Waterloo, ON, Canada N2L 3G1}
\affiliation{Waterloo Institute for Nanotechnology, University of Waterloo, Waterloo, ON, Canada N2L 3G1}

\author{Kouichi Semba}
\affiliation{Advanced ICT Research Institute, National Institute of Information and Communications Technology (NICT), 4-2-1, Nukui-Kitamachi, Koganei, Tokyo 184-8795, Japan}
\affiliation{Present address: Institute for Photon Science and Technology, The University of Tokyo, Tokyo 113-0033, Japan}

\date{\today}

\begin{abstract}
We consider the implementation of two-qubit gates when the physical systems used to realize the qubits possess additional quantum states in the accessible energy range. We use optimal control theory to determine the maximum achievable gate speed for two-qubit gates in the qubit subspace of the many-level Hilbert space, and we analyze the effect of the additional quantum states on the gate speed. We identify two competing mechanisms. On one hand, higher energy levels are generally more strongly coupled to each other. Under suitable conditions, this stronger coupling can be utilized to make two-qubit gates significantly faster than the reference value based on simple qubits. On the other hand, a weak anharmonicity constrains the speed at which the system can be adequately controlled: according to the intuitive picture, faster operations require stronger control fields, which are more likely to excite higher levels in a weakly anharmonic system, which in turn leads to faster decoherence and uncontrolled leakage outside the qubit space. In order to account for this constraint, we modify the pulse optimization algorithm to avoid pulses that lead to appreciable population of the higher levels. In this case we find that the presence of the higher levels can lead to a significant reduction in the maximum achievable gate speed. We also compare the optimal-control gate speeds with those obtained using the cross-resonance/selective-darkening gate protocol. We find that the latter, with some parameter optimization, can be used to achieve a relatively fast implementation of the CNOT gate. These results can help the search for optimized gate implementations in realistic quantum computing architectures, such as those based on superconducting circuits. They also provide guidelines for desirable conditions on anharmonicity that would allow optimal utilization of the higher levels to achieve fast quantum gates.
\end{abstract}

\maketitle

\section{Introduction}
\label{Sec:Introduction}

Over the past two decades, superconducting qubits have made remarkable progress towards the goal of constructing a large quantum computer \cite{Ladd,Buluta,Kjaergaard}. As the superconducting qubit technology matures, it becomes increasingly important to optimize the various aspects of their operation, such as their coherence times and gate speeds. In particular, the question of speed limits can be expressed as follows: what is the minimum amount of time needed to implement a given quantum gate with the minimum required fidelity, e.g.~99.9\%, in a given setup? There have been several studies on this topic for multi-qubit systems, considering different scenarios as it relates for example to the nature of the control parameters and qubit-qubit interactions \cite{AshhabSpeedLimits}. As a general rule, the speed limit for a given two-qubit gate is determined by the two-qubit coupling strength, with a linear proportionality relation between coupling strength and gate speed.

Some superconducting qubit designs that have long coherence times have weak anharmonicities. In other words, the device used to realize the qubit has several quantum states with comparable transition frequencies between them. In the case of particularly weak anharmonicity, the lowest few energy levels are almost equally spaced, and the device behaves almost as a harmonic oscillator. Hence, if one applies a drive signal that is tuned to resonance with the transition between the two lowest energy levels, which are used to encode the qubit states, one must consider the possibility that the same drive signal will induce unwanted near-resonant transitions from the qubit states to the higher levels. A careful analysis of the quantum computer operation must therefore include the higher levels in the modeling of the physical device. Our goal in this work is to investigate the effect of these higher levels on the maximum achievable speed of two-qubit gates.

It is worth mentioning here that weak anharmonicity is not a new development in the field of superconducting qubits. The phase qubit \cite{Martinis}, which is one of the earliest superconducting qubit designs, is also weakly anharmonic. Another point to note is that although the higher energy levels of weakly anharmonic qubits are generally thought of as being detrimental to the operation of the qubit, e.g.~because of the possibility of unintentionally driving the device outside the space of qubit states \cite{Motzoi}, the higher levels can be utilized to enable or enhance certain qubit operations \cite{Strauch,Matsuo,Neeley,DiCarlo,Solenov2014,Solenov2016,Nesterov}.

In order to determine the speed limits for two-qubit gates, we use optimal control theory (OCT), which is a powerful tool to find optimized pulses that can effect various quantum computing tasks \cite{Werschnik}, such as two-qubit and multi-level-system control \cite{Spoerl,Mueller,Reich,Huang,Watts,Heeres,Hu,Wu,Zong}. In particular, by varying the pulse time and observing changes in the achievable gate fidelity, one can use OCT to find the minimum time needed to implement a given quantum gate, e.g.~the CNOT gate, with a given target fidelity. When dealing with simple qubits, e.g.~physical systems for which two qubit states can be identified and manipulated with negligible leakage to other quantum states, OCT algorithms can be applied directly and produce accurate results for the speed limits on quantum gates.

Weakly anharmonic qubits have two properties that complicate the application of standard OCT algorithms. First, if we focus on performing information processing in the qubit space, any unitary operator that implements the desired operation in the qubit space is equally acceptable, regardless of how the additional quantum states are transformed. In other words, there are infinitely many unitary operators that qualify as equally valid choices for the target quantum gate. Another complication is that higher levels of weakly anhamonic qubits are commonly more prone to decoherence and further leakage to other quantum states. As a result, even if these states are present and can be occupied at intermediate times during the implementation of the quantum gate, it can be desirable to avoid populating them as much as possible. The weaker the anharmonicity, the more important this consideration becomes. In this work we implement a modified version of an OCT algorithm with adjustments designed to deal with these two complications.

The remainder of this paper is organized as follows: In Sec.~\ref{Sec:Setup} we introduce the physical setup of two coupled weakly anharmonic qubits. In Sec.~\ref{Sec:Algorithm} we describe the OCT algorithm used in this work. In Sec.~\ref{Sec:Parameters} we give the parameters used in our numerical calculations. In Sec.~\ref{Sec:Results} we present the results of our numerical OCT calculations. For comparison we present gate time and fidelity results for the cross-resonance/selective-darkening (CR/SD) protocol in Sec.~\ref{Sec:CRSD}. Section \ref{Sec:Conclusion} contains concluding remarks.

\section{Two coupled weakly anharmonic qubits}
\label{Sec:Setup}

We consider a system composed of two coupled qubits. With superconducting qubits in mind, we think of each qubit as being a multi-level quantum system (which can also be called a qudit) and the lowest two energy levels are used to encode the qubit states $\ket{0}$ and $\ket{1}$.

The Hamiltonian of a driven multi-level system whose lowest two levels are used as a qubit can often be expressed as:
\begin{equation}
%
\hat{H} = \sum_{j=0}^{N-1} \omega_j \hat{\Pi}_j + \sum_{j=1}^{N-1} \epsilon(t) \lambda_j \left( \hat{\sigma}_j^+ + \hat{\sigma}_j^- \right),
\label{Eq:SingleQubitHamiltonian}
\end{equation}
where the index $j$ enumerates the $N$ energy eigenstates of the multi-level system (evaluated in the absence of driving) that are kept in the theoretical model (with the ground state labeled by the index $j=0$), $\omega_j$ are the energies of the different states (and we shall set $\omega_0=0$), $\hat{\Pi}_j$ are the projectors for the different states $j$ ($\hat{\Pi}_j=\ket{j}\bra{j}$), $\epsilon(t)$ is the time-dependent amplitude of the driving field, $\lambda_j$ are coefficients that set the relation between the driving-induced coupling matrix elements of the different transitions, $\hat{\sigma}_j^+=\ket{j}\bra{j-1}$ and $\hat{\sigma}_j^-=\ket{j-1}\bra{j}$. It should be noted that the most important piece of information about $\lambda_j$ is the relation between the different coefficients, i.e.~not each coefficient separately, because they are all multiplied by the common driving field amplitude $\epsilon(t)$. Throughout this work, we shall use the same units for energy and frequency, i.e.~we set $\hbar=1$.

Although our interest and results will not be limited to the case of extremely weak anharmonicity, we use a model of a truncated weakly harmonic oscillator for our calculations.  In particular, we set $\lambda_j = \sqrt{j}$ as an approximation. This behavior is exact for a harmonic oscillator. Experimental results show that it remains a good approximation for weakly anharmonic superconducting qubit devices such as the phase qubit \cite{Martinis} and transmon \cite{Koch}. With the approximation of near-harmonicity, we also ignore direct coupling between states $\ket{j}$ and $\ket{j\pm m}$ with $m\neq 1$. We emphasize that even when the devices deviate substantially from the harmonic oscillator approximation, general relations such as the increase of $\lambda_j$ with increasing $j$ tend to remain valid. This property is related to the fact that the extension of wave functions generally increases as we go to higher energy levels. With increasing extension of the multi-level system's wave functions, the coupling to external fields becomes stronger, which corresponds to increasing values of $\lambda_j$.

Using the harmonic oscillator relations described above, we can simplify the notation by defining the operator
\begin{eqnarray}
\hat{a} & = & \sum_{j=1}^{N-1} \sqrt{j} \hat{\sigma}_j^- \nonumber \\
\hat{a}^{\dagger} & = & \sum_{j=1}^{N-1} \sqrt{j} \hat{\sigma}_j^+.
\label{Eq:LadderOperators}
\end{eqnarray}
These operators are the harmonic oscillator annihilation and creation operators truncated to the lowest $N$ energy levels.

In this work, we shall not make the rotating wave approximation, which could speed up our calculations but would also ignore the so-called counter-rotating terms in the Hamiltonian. These terms could lead to some non-negligible effects when dealing with high-bandwidth or high-power driving fields. We wish our calculations to capture any such effects if they arise in the dynamics resulting from the optimized pulses that we obtain in the calculations.

In present-day designs of superconducting qubits, since the coupling between neighboring qubits is typically mediated by mechanisms similar to those that describe the coupling to external driving fields, we expect the same operators to appear in the driving and coupling terms in the Hamiltonian. As a result, we approximate the two-qubit coupling term in the Hamiltonian by
\begin{equation}
\hat{H}_C = g \left( \hat{a}_1 + \hat{a}_1^{\dagger} \right) \otimes \left( \hat{a}_2 + \hat{a}_2^{\dagger} \right),
\label{Eq:CouplingTerm}
\end{equation}
where $g$ is the coupling strength and the subscripts ``1'' and ``2'' label the two qubits. In this work we assume that $g$ is fixed, which is a common situation in experiment.

To help separate different phenomena that can be at play in the system under consideration, we shall take $g$ to be much smaller than the anharmonicity parameters (which are defined as $\eta_j=\omega_j-j\omega_1$), and we take the anharmonicity parameters to be much smaller than the single-qubit Larmor frequencies, i.e.~the frequency $\omega_1$ for each qubit. The detuning between the two qubits $\omega_1^{(1)}-\omega_1^{(2)}$ is typically designed to be much smaller than $\omega_1^{(1)}$ and $\omega_1^{(2)}$ but much larger than $g$, because a large detuning leads to slower two-qubit gates in practical setups, and a small detuning leads to frequency crowding. Since the Larmor frequencies and coupling strengths are typically separated by only two orders of magnitude, the inter-qubit detuning and the anharmonicity will be at the same scale. This ordering of energy scales does in fact correspond to commonly used systems of superconducting qubits \cite{Chow,Barends}. The coupling strength is usually not much smaller than the anharmonicity. However, we expect that this fact does not affect our main conclusions. Furthermore, the interpretation of our numerical results can be straightforwardly applied to more general situations, such as the case of qubits with strong anharmonicity.

Another point worth noting here is that the coupling term mixes the computational basis states, e.g.~$\ket{01}$ and $\ket{10}$, causing the energy eigenstates to be superpositions of these states. One might intuitively think that this mixing will reduce the fidelity of any operation. However, in practice all operations are performed in the basis of energy eigenstates, and the small perturbations in these states caused by the coupling term are naturally absorbed into the definition of the computational basis states. As such, these perturbations do not in themselves constitute an error in any given protocol. In fact, the mixing in the energy eigenstates can be seen as the mechanism that enables certain two-qubit gate protocols \cite{DeGroot2010}.

In the absence of higher levels, and assuming that there are no constraints on the control fields $\epsilon(t)$, there are several standard methods for implementing various two-qubit quantum gates, such as the CNOT gate, which is described by the unitary operator
\begin{equation}
U_{\rm CNOT} = \left( \begin{array}{cccc} 1 & 0 & 0 & 0 \\ 0 & 1 & 0 & 0 \\ 0 & 0 & 0 & 1 \\ 0 & 0 & 1 & 0 \end{array} \right)
\label{Eq:UCNOT}
\end{equation}
in the basis $\left\{ \ket{00}, \ket{01}, \ket{10}, \ket{11} \right\}$, where the first and second indices represent the states of the control and target qubits, respectively. These methods can in general be applied to weakly anharmonic qubits, although additional care needs to be taken to deal with the higher energy levels. One such method is the CR/SD gate \cite{Rigetti,DeGroot2010}, which we shall use for comparison with some of our OCT results below.

In the absence of higher levels, the minimum time required to perform the CNOT gate is $T_0=\pi/(4g)$, as explained in Ref.~\cite{AshhabSpeedLimits}. This gate time can be achieved by setting the qubit bias fields $\epsilon(t)$ to large values to implement a controlled phase gate. With the proper phase value, the controlled phase gate is equivalent to the CNOT gate, up to single-qubit rotations. This speed limit can also be approached when using strong driving with ac-based implementations such as the CR/SD gate. For definiteness, we shall focus on the CNOT gate in our analysis below and compare the minimum gate time for weakly anharmonic qubits with the $\pi/(4g)$ time mentioned above.

In experimental realizations of quantum gates, one usually expects a small amount of error to remain even after optimization. These errors are typically a result of imperfections in the experimental implementation. We note that in OCT calculations with a sufficiently large number of adjustable parameters, as is the case in our zero-loss OCT calculations, there exist pulses that lead to perfect gate implementations in theory. We shall therefore not analyze small residual errors in relation to OCT calculations. Besides, these small errors are unrelated to the question of the speed limits that is the main topic of this work.

\section{Pulse optimization algorithm}
\label{Sec:Algorithm}

We use numerical OCT techniques to search for control pulses that effect a given target unitary operator. As mentioned above, we use the CNOT gate as a representative target gate in our analysis of two-qubit gates. We expect that our general conclusions regarding the role of the higher levels in speeding up or slowing down two-qubit gates is not specific to our choice of the CNOT gate. An alternative approach is to leave the target gate unspecified and instead let the optimization algorithm accept any perfect entangler to maximize the fidelity for a given set of system parameters, as explained in Ref.~\cite{Watts}. As our pulse search method we use the gradient ascent pulse engineering (GRAPE) algorithm \cite{Khaneja}, which has the advantage of being fast even for large numbers of control parameters.

In OCT algorithms for finding the optimal pulse for implementing a unitary operator, the goal typically is to maximize the fidelity
\begin{equation}
F = \left| \frac{ {\rm Tr} \left\{ U_{\rm Target}^{\dagger} U(T) \right\}}{d} \right|^2,
\label{Eq:Fidelity}
\end{equation}
where $U_{\rm Target}$ is the desired target operator, $U(T)$ is the candidate operator that is implemented by the numerically calculated pulse (which is updated and improved by the algorithm), and the factor $d$ in the denominator is the dimension of the Hilbert space ($d=4$ for a two-qubit system). The fidelity $F$ quantifies the overlap between $U(T)$ and $U_{\rm Target}$. When these two operators are identical, Eq.~(\ref{Eq:Fidelity}) gives $F=1$. When dealing with weakly anharmonic qubits and expanded Hilbert spaces, we shall define the operators in such a way that the maximum value of the fidelity is $F=1$ in this case as well.

In the GRAPE algorithm \cite{Khaneja}, the pulse is assumed to be piecewise constant with the total pulse time divided into $N$ time steps. The operator $U(T)$ is therefore given by
\begin{equation}
U(T) = U_N U_{N-1} \cdots U_2 U_1,
\end{equation}
where $U_j$ is the unitary operator that describes the evolution in the $j$th time step:
\begin{equation}
U_j = \exp \left\{ -i \Delta t \left( \hat{H}_0 + \sum_{k=1}^{m} u_k(j) \hat{H}_k \right) \right\}
\end{equation}
$\Delta t$ is the duration of the time step, $\hat{H}_0$ is the fixed part of the Hamiltonian, $m$ is the number of control parameters, $u_k(j)$ is the value of the $k$th control parameter in the $j$th time step, and $\hat{H}_k$ is the $k$th control Hamiltonian. In each iteration of the optimization procedure, the calculation proceeds by first calculating the operators
\begin{equation}
X_j = U_j \cdots U_2 U_1
\label{Eq:GRAPEX}
\end{equation}
and
\begin{equation}
P_j = U_{j+1}^{\dagger} \cdots U_{N-1}^{\dagger} U_N^{\dagger} U_{\rm Target}
\label{Eq:GRAPEP}
\end{equation}
for all values of $j$. The pulse update is then determined by calculating the derivatives of the fidelity with respect to the different control parameters $u_k(j)$ in the different time steps $j$
\begin{equation}
\frac{dF}{du_k(j)} = -\frac{1}{8} {\rm Re} \left[ i \Delta t {\rm Tr} \left\{ P_j^{\dagger} \hat{H}_k X_j \right\} {\rm Tr} \left\{ X_j^{\dagger} P_j \right\} \right].
\label{Eq:FidelityDerivative}
\end{equation}
With this information at hand, one can update the control parameters $u_k(j)$ by moving along the direction of the gradient of $F$ to maximize the fidelity improvement in each iteration. Importantly, it has been shown that for typical control problems there are no local maxima in $F$ that could prevent the algorithm from finding the absolute maximum \cite{Brif}.

We now consider what modifications we need to make in order to apply the GRAPE algorithm to multi-level systems that contain the qubit space as well as additional quantum states outside the computational space. First we consider the fact that we are looking for a certain unitary operator in the qubit space, regardless of what transformation is effected in the remainder of the Hilbert space. This situation means that the fidelity should be the same for all the equivalent operators that differ only in their effect on initial states outside the qubit space. In other words, there should not be any cost associated with the part of $U(T)$ that describes the transformation of states in the irrelevant subspace of the Hilbert space. This goal can be achieved by using as the target operator a matrix that has the desired matrix elements in the relevant subspace and zero matrix elements for the rest of the matrix. Such a matrix would not be a unitary operator. It just serves the purpose of guiding the search to the space of acceptable (and equivalent) target operators without favoring any member of this set over any other. Note that a special case of this technique was used in the study of single-qubit optimal control in a three-level quantum system \cite{Rebentrost}.

The other consideration that we would like to incorporate into the algorithm is the desire to avoid going too high in the energy level ladder outside the qubit space, even at intermediate times during the dynamics. This condition is motivated by the fact that these higher energy levels tend to be associated with increased dissipation and can be prone to further leakage that causes information loss. We therefore need to include some penalty for populating these states during the dynamics. This goal can be achieved by introducing a loss factor that shrinks the matrix elements in the dynamical evolution operator that correspond to the higher levels. Such a factor can be included by replacing the operator product in the fidelity (Eq.~\ref{Eq:Fidelity}) by
\begin{equation}
U_{\rm Target}^{\dagger} L U_N L U_{N-1} \cdots L U_2 L U_1,
\end{equation}
where
\begin{eqnarray}
L & = & \exp \left\{ -\Gamma \Delta t \right\},
\\
\Gamma & = & \left( \begin{array}{ccccc}
\gamma_1 & 0 & 0 & \cdots & 0 \\
0 & \gamma_2 & 0 & & 0 \\
0 & 0 & \gamma_3 & & 0 \\
\vdots & & & \ddots & \vdots\\
0 & 0 & 0 & \cdots & \gamma_M \end{array} \right),
\end{eqnarray}
$\gamma_l$ are loss rates for the different quantum states. Note that although we are using this loss factor as a computational tool to steer the optimization algorithm away from solutions that involve occupying certain quantum states, this loss model does in fact have a physical meaning. It can describe a real dissipative loss of probability from the physical subspace under study, and it corresponds to non-hermitian stochastic dynamics \cite{Molmer}. A similar modeling of loss in an OCT problem was used in Ref.~\cite{Goerz}. In our optimization algorithm, populating the undesired states would lead to a reduction in the corresponding matrix elements in the evolution operator, which would propagate to other matrix elements and lead to a reduced fidelity at the final time. As a result, the search algorithm moves away from such situations and towards pulses that keep the system as much as possible in the non-decaying subspace. To implement this change in the GRAPE algorithm, we replace the definitions in Eqs.~(\ref{Eq:GRAPEX}) and (\ref{Eq:GRAPEP}) by:
\begin{eqnarray}
X_j & = & L U_j \cdots L U_2 L U_1
\nonumber \\
P_j & = & U_{j+1}^{\dagger} L \cdots U_{N-1}^{\dagger} L U_N^{\dagger} L U_{\rm Target}.
\end{eqnarray}
With this modification we can use Eq.~(\ref{Eq:FidelityDerivative}) to update the control parameters $u_k(j)$ and hence optimize the control pulses with this additional consideration incorporated into the algorithm. We note here that this modification to the algorithm can be used to suppress occupying any state, possibly for reasons other than decoherence. We also note that one could alternatively add to the algorithm specific decoherence terms, e.g.~in Lindblad form. However, the proper description of decoherence in an open quantum system would require us to work with objects that are more general than unitary operators, e.g.~completely positive maps, which would somewhat complicate the calculations without any benefit for our purposes, and we do not do so here.

In the scenario that we analyze in this work, the fixed part of the Hamiltonian is
\begin{equation}
\hat{H}_0 = \sum_{k=1}^{2} \sum_{j=1}^{N-1} \omega_j^{(k)} \hat{\Pi}_j^{(k)} + g \left( \hat{a}_1 + \hat{a}_1^{\dagger} \right) \otimes \left( \hat{a}_2 + \hat{a}_2^{\dagger} \right),
\end{equation}
and there are two control Hamiltonians,
\begin{equation}
\hat{H}_k = \hat{a}_k + \hat{a}_k^{\dagger},
\end{equation}
with $k=1,2$, and the control parameters $u_k(j)$ are the values of the two drive fields $\epsilon_k(t)$ in the $N$ time steps.

\section{Calculation parameters}
\label{Sec:Parameters}

In presenting the results below, we shall use the first qubit's Larmor frequency $\omega_1^{(1)}$ as the reference energy. In other words, all the energies and frequencies below are given in dimensionless form and should be understood as being divided by this energy unit. The time unit is accordingly $2\pi/\omega_1^{(1)}$. We shall, however, present most of the results relative to the minimum CNOT gate time for two simple qubits, namely $T_0=\pi/(4g)$. In all the calculations, we set the second qubit's Larmor frequency to $\omega_1^{(2)}=0.9$. The coupling strength is set to $g=0.0025$.

The anharmonicities $\eta_j^{(k)}=\omega_j^{(k)}-j\omega_1^{(k)}$ are set to $\eta_2^{(1)} = \eta_2^{(2)} = -0.11$, $\eta_3^{(1)} = \eta_3^{(2)} = -0.19$ and $\eta_4^{(1)} = \eta_4^{(2)} = -0.28$, unless otherwise stated. We choose negative values with magnitudes that grow as we go up the energy level ladder in accordance with the basic behavior of phase qubit and transmon energy level structures. The anharmonicity of a phase qubit or transmon grows as $\eta_{j}=(j^2-j)\times\eta_2/2$ to lowest order, which gives $\eta_3=3\eta_2$ and $\eta_4=6\eta_2$. We do not use this formula, because it would make $\eta_j^{(k)}$ on the order of $\omega_1^{(k)}$ for $j=4$, meaning that the formula cannot be a good approximation for $j\geq 4$. As we shall discuss in Sec.~\ref{Sec:Results}, there are intuitive explanations for various aspects of our results, and the exact functional dependence of $\eta_j$ on $j$ does not seriously complicate the interpretation of our results. Note also that the value $|\eta_2^{(k)}|=0.11$ is close to the detuning between the two qubits ($\omega_1^{(1)}-\omega_1^{(2)}=0.1$). By varying the anharmonicity, we shall see in the next two sections that this choice does not seem to have a noticeable effect on the OCT results, but it can drastically affect the CR/SD protocol.

The relaxation rates of nearly harmonic superconducting qudits grow approximately as $\Gamma_{j\rightarrow j-1}=j\Gamma_{1\rightarrow 0}$, where $\Gamma_{j\rightarrow j-1}$ is the relaxation rate from the state $\ket{j}$ to the state $\ket{j-1}$. It might therefore seem logical to set the loss rates $\gamma_j = j \gamma_1$. However, since our focus is on the question of how higher levels affect speed limits, and we use the loss rates $\gamma_j$ as a computational tool rather than to simulate a real relaxation process, we do not use the formula for $\Gamma_{j\rightarrow j-1}$ to set the values of $\gamma_j$. Instead, we set $\gamma_1=0$ and use various combinations of values for the loss rates of the higher levels.

The number of time steps was set to $N=10^3$ for pulse times up to $50$, $N=10^4$ for pulse times between $50$ and $400$, and $N=2\times10^4$ for pulse times longer than $400$. These parameters were chosen to balance between the desire to keep the computation time relatively short and making sure that the results do not change if we increase the number of time steps. We verified that with the above parameters the number of time steps is large enough to make further increases unnecessary. For all of the simulations, $10^4$ iterations were used.

For the initial guess for the driving fields, we used randomly generated, and therefore highly noisy, signals taken from four different distributions. These distributions are all uniform, i.e.~they produce uniformly distributed random numbers for each one of the parameters $u_k(j)$. The four distributions are defined by two criteria: the width is either 1 or 10, and the distribution either is centered or starts at 0. Each calculation with the parameters described above takes about one day on a single core of a personal computer.

\section{Results}
\label{Sec:Results}

The presence of additional quantum states outside the qubit space opens new channels for the system dynamics. This modification to the system brings both positive and negative aspects. On one hand, the new channels for the dynamics allow new possibilities that could be useful if one can find ways to harness them. On the other hand, they create new channels for leakage of the quantum state outside the qubit space, which leads to deviations from the intended dynamics, unless care is taken to suppress the leakage. We shall see in this section that the additional states can speed up or slow down the implementation of two-qubit gates, depending on the details of the situation being considered.

First we perform calculations in the case where we do not assign any penalty to occupying higher levels. In other words we treat them as harmless additional energy levels, as long as one makes sure to return to the qubit space at the end of the controlled operation. Even with no explicitly assigned penalty for occupying the higher levels, it is not obvious that refocusing the quantum state back into the qubit space is a simple task. One might therefore expect that the mere presence of the additional quantum states could lead to a reduced gate speed. We shall see shortly that in general the opposite is true. It should also be emphasized that we start our analysis with the implicit assumption that the truncated oscillator model provides a good approximation for the system. Our results will soon show that we must remain cautious about this assumption when dealing with weakly anharmonic qubits.

\begin{figure}[h]
\includegraphics[width=8.0cm]{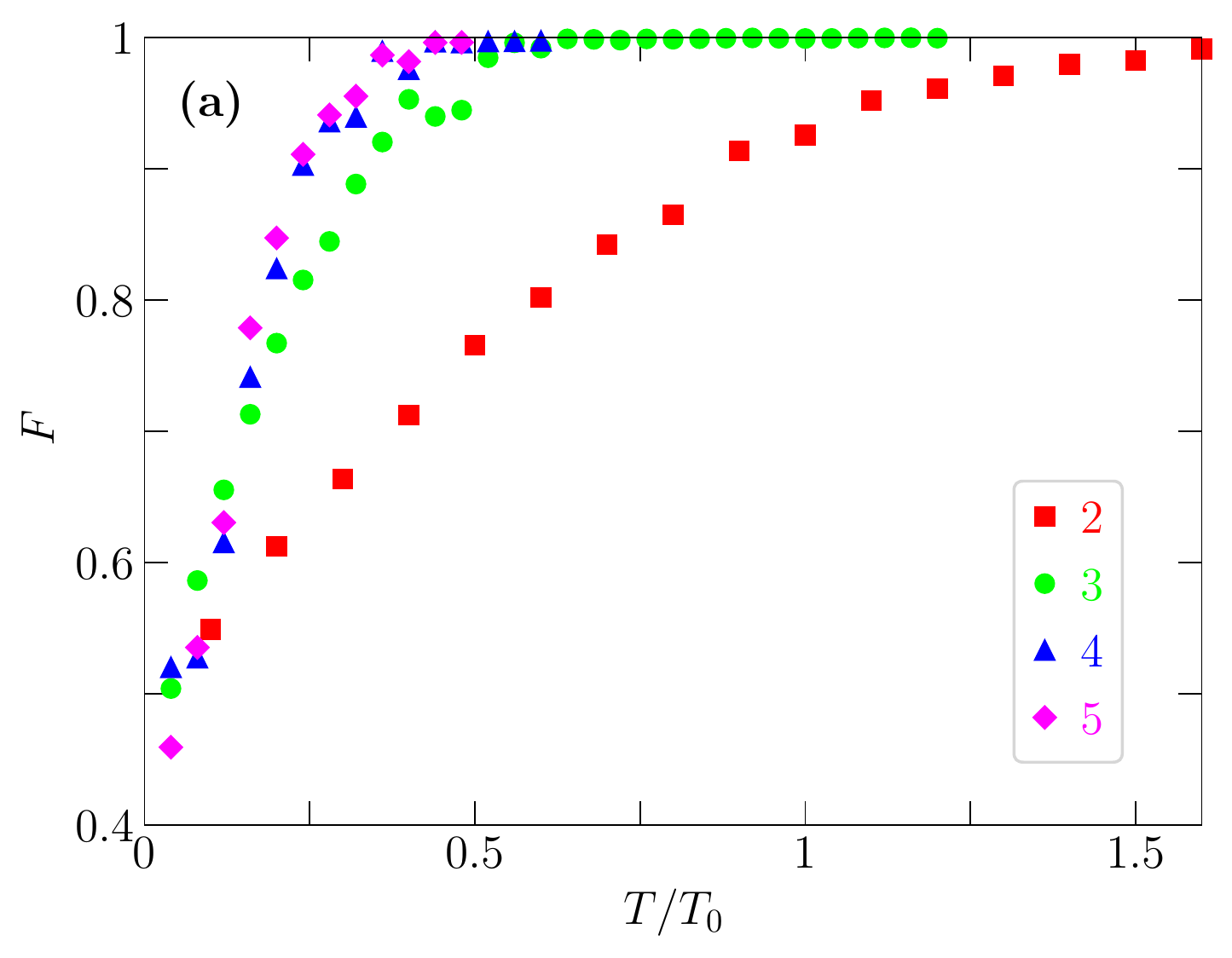}
\includegraphics[width=8.0cm]{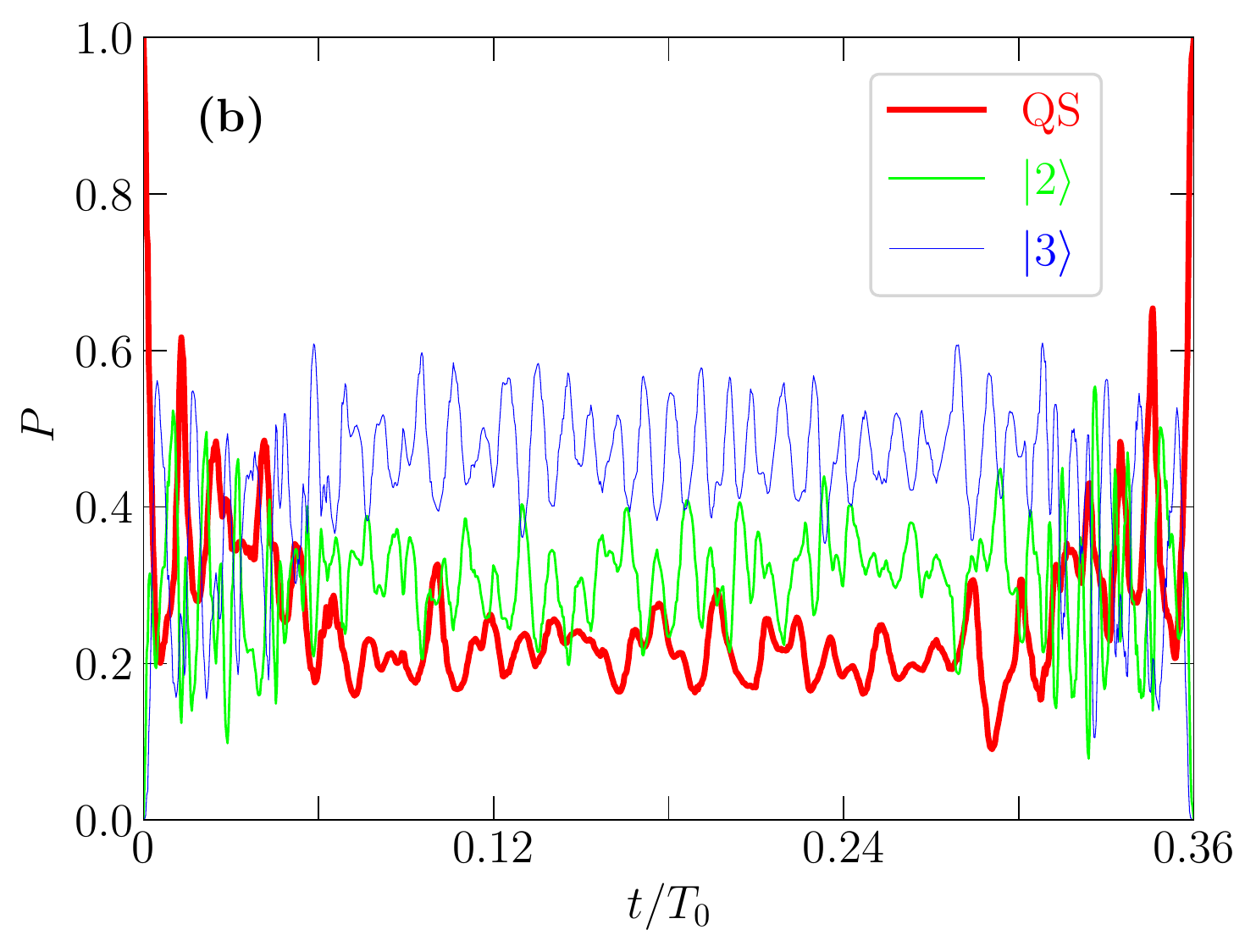}
\caption{(a) The fidelity $F$ of the CNOT gate implemented with the optimal numerically obtained pulse as a function of the allowed time $T$ [measured in units of the minimum CNOT gate time for simple qubits $T_0=\pi/(4g)$] in the absence of a loss factor in the simulations. The red squares, green circles, blue triangles and magenta diamonds correspond, respectively, to having a total of 2, 3, 4 or 5 energy levels for each qubit. The system parameters are described in Sec.~\ref{Sec:Parameters}. (b) Populations $P$ in the different subspaces as functions of time for the blue triangle at $T/T_0=0.36$ in Panel (a). The populations are averaged over the 4 computational basis states, i.e.~$\{\ket{00}, \ket{01}, \ket{10}, \ket{11}\}$, as initial states. The thick red line shows the probability to be in the qubit space (QS). The medium-width green line corresponds to having at least one qubit in the state $\ket{2}$ but no qubits in the state $\ket{3}$. The thin blue line corresponds to having at least one qubit in the state $\ket{3}$.}
\label{Fig:OCTZeroLoss}
\end{figure}

The CNOT gate fidelity as a function of allowed pulse time for different total numbers of energy levels is shown in Fig.~\ref{Fig:OCTZeroLoss}(a). Perhaps counter-intuitively, the presence of the higher levels allows faster implementations of the gate. Taking into consideration that the minimum gate time for the case of simple qubits is $T_0$, the addition of a third level per qubit reduces the minimum gate time by a factor of 2. Increasing the number of additional levels that we include in the simulations leads to faster gates. This trend slows down and is barely visible when comparing the four- and five-level simulations. The speedup with increasing number of levels can be understood by noting that the bottleneck for the two-qubit gate speed is the photon-exchange dynamics induced by the coupling term in the Hamiltonian. Opening extra channels for the photon exchange can speed up the completion of the two-qubit gate operation. Importantly, because the coupling strength increases as we go to higher energy levels, it is in fact advantageous to let the photon exchange occur through the higher levels. One mechanism that can stop the increase in gate speed with increasing Hilbert space size is the time needed to excite the system from the qubit space to the high levels and bring them back to the qubit space at the end. Studies on this process have shown that it requires a minimum pulse time that scales inversely with the anharmonicity, with a power close to one \cite{Khani,Zhu}. With our parameters, i.e.~$\eta_j^{k}\sim 0.1 \omega_1^{(k)}$, this mechanism would become a limiting factor when the number of states per qubit is around 5, which could partly explain the stagnation in the speedup seen in Fig.~\ref{Fig:OCTZeroLoss}.

To illustrate the role of the higher levels in the gate dynamics, we take the optimized pulse for one of the data points in Fig.~\ref{Fig:OCTZeroLoss}(a) and plot the populations of different parts of the Hilbert space as functions of time Fig.~\ref{Fig:OCTZeroLoss}(b). It is clear that the population leaves the qubit space during the gate dynamics but returns to the qubit space at the final time. To avoid crowding the figure, we have averaged the population results over four initial states corresponding to the four computational basis states in the qubit space. For the initial states $\ket{00}$ and $\ket{01}$ about 50\% of the population remains in the qubit space at intermediate times, while for the initial states $\ket{10}$ and $\ket{11}$ the population in the qubit space is below 10\% for most of the pulse duration. The highest energy level has the highest average population at intermediate times. We note here that there is a reason why we chose a data point whose $T$ value is just below the minimum gate time in Fig.~\ref{Fig:OCTZeroLoss}(a). In particular, we avoid longer pulse times, because long pulse times allow an infinite number of pulses that all lead to fidelity values above any threshold that we set. In other words, if we use a long pulse time, we can take a long and winding path in unitary-operator space and still reach the target gate at the final time. The search algorithm does not favor any path over any other, as long as they reach the target gate at the final time. We can then expect to obtain more irregular dynamics compared to those shown in Fig.~\ref{Fig:OCTZeroLoss}(b). In contrast, for pulse times below the minimum gate time we expect that there will be a unique optimal pulse that utilizes the available quantum states in an optimal manner.

We performed similar calculations for the fidelity as a function of pulse time with the anharmonicities all set to zero. The results remained unchanged to within the margin of computational fluctuations. This result might seem paradoxical; zero anharmonicity suggests that the qubits become harmonic oscillators, and it should be impossible to perform inherently quantum operations, such as the CNOT gate, on harmonic oscillators with bilinear coupling. However, the truncation of the Hilbert space creates an anharmonicity and makes it possible to perform quantum operations on the system.

The above results demonstrate the limitations of the truncated oscillator approximation and the importance of including a penalty term in the cost function to avoid pulses that lead to populating higher energy levels. In the absence of any such term and considering weakly anharmonic qubits with a large number of extra quantum states from which we keep a few levels in the OCT calculation, the calculation will generally produce the result that the fastest two-qubit gate implementation involves exciting the device to the highest levels at intermediate times. However, if we keep $N$ levels in the theoretical approximation of a weakly anharmonic qubit and find that the optimal pulse drives the population up to the highest level during the gate dynamics, this result would be an indication that the approximation (i.e.~the truncation of the energy levels) was not justified. The approximation is only justified if the populations of the highest levels that are kept in the theoretical model remain small, meaning that the ignored levels would have even smaller populations and ideally a negligible effect on the final results. Instead, if we find that the $N$th level is significantly populated, we have to keep the $(N+1)$th level in the theoretical model as well. These complications do not arise if the anharmonicity is large enough that individual control of each qudit's transitions can be performed significantly faster than the two-qubit gate time. However, for anharmonicitiy values that are so small that full single-qudit control cannot be performed significantly faster than the CNOT gate time $T_0$, our results based on the truncated model become suspect. One apparent solution to this difficulty is to use a more accurate model with a large number of energy levels per qudit. However, with present-day superconducting circuits, one cannot realistically utilize more than a few energy levels without having serious detrimental effects of dissipation. Even if future devices have good coherence properties extending to many excited states, one must still worry about uncontrolled leakage to very high levels if one attempts fast control with weak anharmonicity. Besides, characterizing and properly modeling higher levels becomes increasingly difficult for a realistic setup. Instead of dealing with these complications, an alternative approach is to avoid pulses that drive the system too high up the energy level ladder. We achieve this goal by adding a penalty to occupying higher levels in our calculations. We emphasize that the role of the penalty term is to make the pulse search algorithm, i.e.~the OCT algorithm, look for pulses that avoid exciting higher levels. Adding this term to the OCT calculation does not necessarily reflect a change in the physical parameters of the system. We also emphasize that the anharmonicity of superconducting qubits generally increases as we go up the energy level ladder. In some cases, the anharmonicity becomes strong after the first few energy levels, which would naturally eliminate the need to worry about uncontrolled leakage to higher levels. We shall come back to this point at the end of this section.

\begin{figure}[h]
\includegraphics[width=8.0cm]{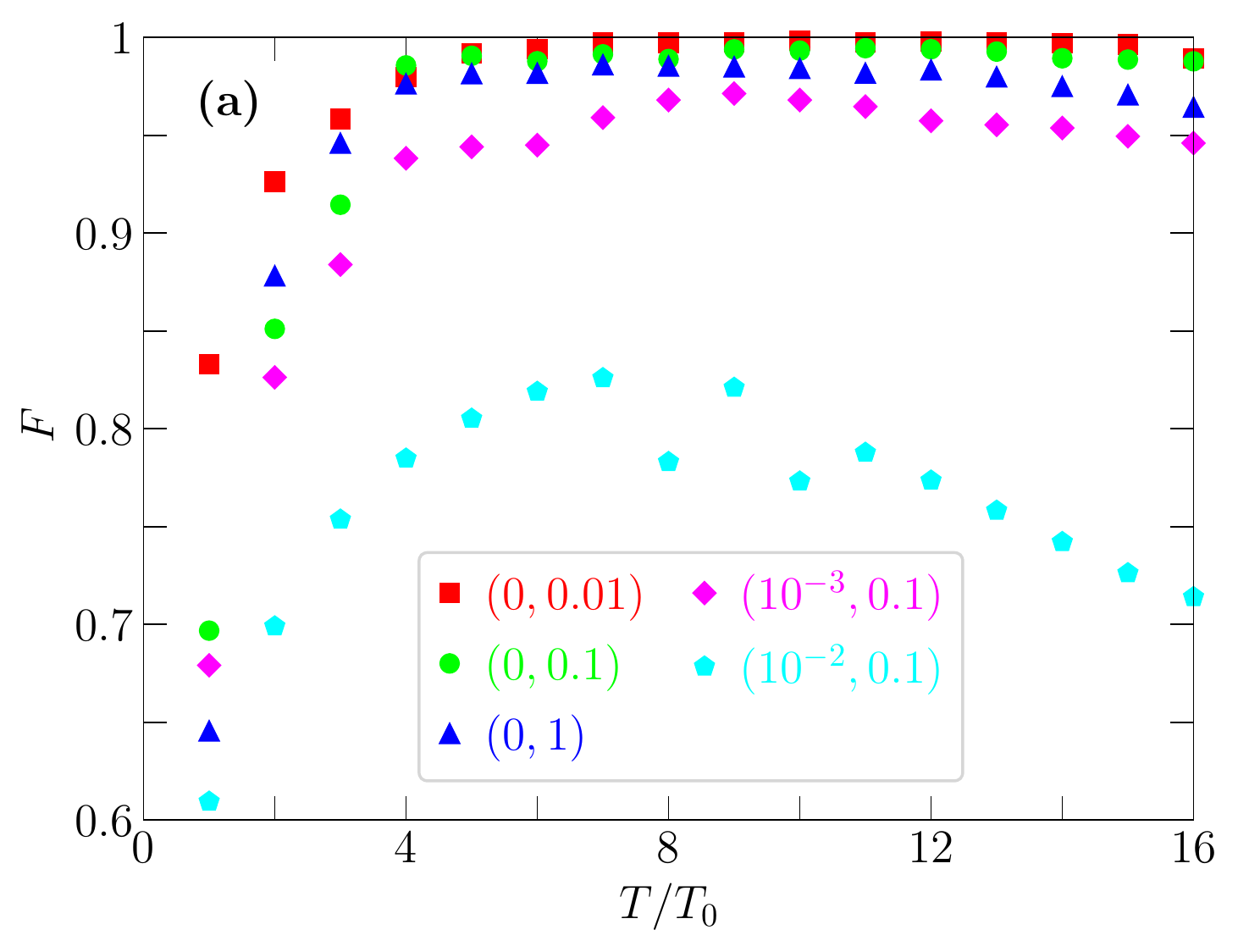}
\includegraphics[width=8.0cm]{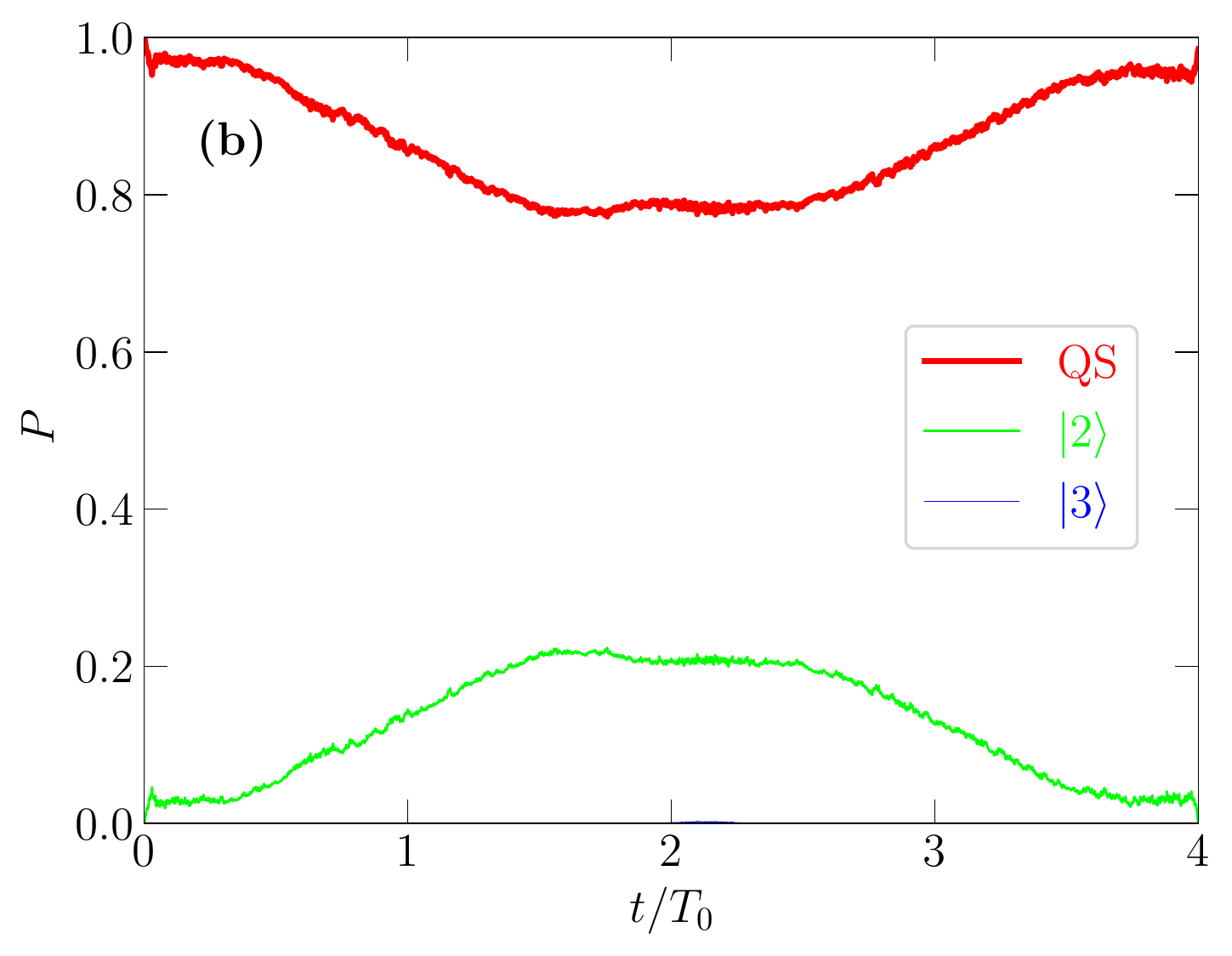}
\caption{(a) Maximum fidelity as a function of pulse time, as in Fig.~\ref{Fig:OCTZeroLoss}(a), but including the loss factor $L$ and varying the loss rates. In all cases, we keep four energy levels per qubit. The red squares, green circles, blue triangles, magenta diamonds and cyan pentagons correspond, respectively, to the loss rate combinations $(\gamma_2,\gamma_3)=(0,10^{-2})$, $(0,10^{-1})$, $(0,1)$, $(10^{-3},10^{-1})$ and $(10^{-2},10^{-1})$. The same values of the loss rates are used for both qubits. All the data sets reach their maximum or saturation values around the same value of $T$. (b) Populations $P$ in the different subspaces as functions of time, as in Fig.~\ref{Fig:OCTZeroLoss}(b), for the green circle at $T/T_0=4$ in Panel (a). The population of the state $\ket{3}$ is very low, making the blue line barely visible.}
\label{Fig:OCTVaryingLossRates}
\end{figure}

We now calculate the fidelity as a function of pulse time with a loss factor added to the simulations. The results for a number of loss rate combinations are shown in Fig.~\ref{Fig:OCTVaryingLossRates}(a). We note here that we set $\gamma_1=0$ in all our simulations, because we are using the loss factor to discourage population of higher energy levels rather than to accurately model a physical dissipation process. In all the data sets in Fig.~\ref{Fig:OCTVaryingLossRates}(a), the fidelity reaches a maximum value, after which it either remains flat or starts to decrease. The main feature that we emphasize in this figure is that the pulse time at which the maximum is first reached is comparable for all data sets, all in the range $4<T/T_0<8$. In particular, the three data sets that have $\gamma_2=0$ all seem to become flat starting around $T/T_0= 4$. The value $T\approx 4T_0$ can therefore be identified as the realistic minimum gate time for this combination of anharmonicity values. It clearly represents a significant slowdown relative to the one that we obtained in the zero-loss calculations ($T\approx 0.4 T_0$). It is also longer than the minimum gate time for simple qubits ($T_0$). This slowdown is what one would intuitively expect based on the consideration that the desire to avoid leakage imposes a constraint on the control signals and limits them to the weak-driving regime. We also note that, especially when $\gamma_2$ is not negligibly small, the maximum fidelity is reduced as a result of adding the loss factor. The higher the loss rates, the lower the maximum fidelity. This effect is at least partly physical, because exciting higher levels can be minimized but not completely eliminated, except in the infinite-time limit. After the fidelity reaches its maximum value, the figure shows a slow decline in fidelity with increasing pulse time. This feature must be a computational artefact. Such a reduction in fidelity can occur if the GRAPE time step $\Delta t$ is not much shorter than the qubits' Larmor periods, i.e.~if $\omega_1^{(1)} \Delta t$ is not much smaller than 1, because piecewise-constant functions cannot approximate resonant driving signals in this case. However, we do not believe that this effect has a significant impact on our data. At the largest value of $T$ in Fig.~\ref{Fig:OCTVaryingLossRates}, $\omega_1^{(1)} \Delta t=25$. To confirm that this number is not too small, we performed additional simulations in which we reduced $N$ in the GRAPE algorithm by a factor of 2. The fidelity decreased by only about 1\% at large $T$ values, which means that we cannot expect a significant increase in fidelity by increasing $N$. As a result, we suspect that the decrease in fidelity is caused mainly by slower convergence for the cases with longer pulse times. We note again that we set $\gamma_1=0$ in all of our simulations. A finite value of $\gamma_1$ would describe loss within the qubit space and would naturally lead to a reduction in fidelity at long times. However, this effect does not occur in our simulations.

To illustrate that the loss factor is serving its intended purpose, we show one example of the gate dynamics in Fig.~\ref{Fig:OCTVaryingLossRates}(b). In contrast to the dynamics in the zero-loss case [Fig.~\ref{Fig:OCTZeroLoss}(b)], now we can see that the population remains mostly in the qubit space and the higher-level population is suppressed. The state $\ket{3}$, which is the only decaying state in this simulation, is almost not populated at all. Here we note that it was not desirable for the plot in Fig.~\ref{Fig:OCTVaryingLossRates}(b) to take a pulse time that is much shorter than the minimum gate time. If we take a very short pulse time, the algorithm might find pulses that populate higher levels even in the presence of the loss factor, as the penalty from occupying the higher levels might be offset by the gain in gate speed when utilizing those higher levels. For this reason and the one described in relation to Fig.~\ref{Fig:OCTZeroLoss}(b), choosing a pulse time that is at the speed limit is ideal for illustrating the different mechanisms at play during the gate dynamics.

\begin{figure}[h]
\includegraphics[width=8.0cm]{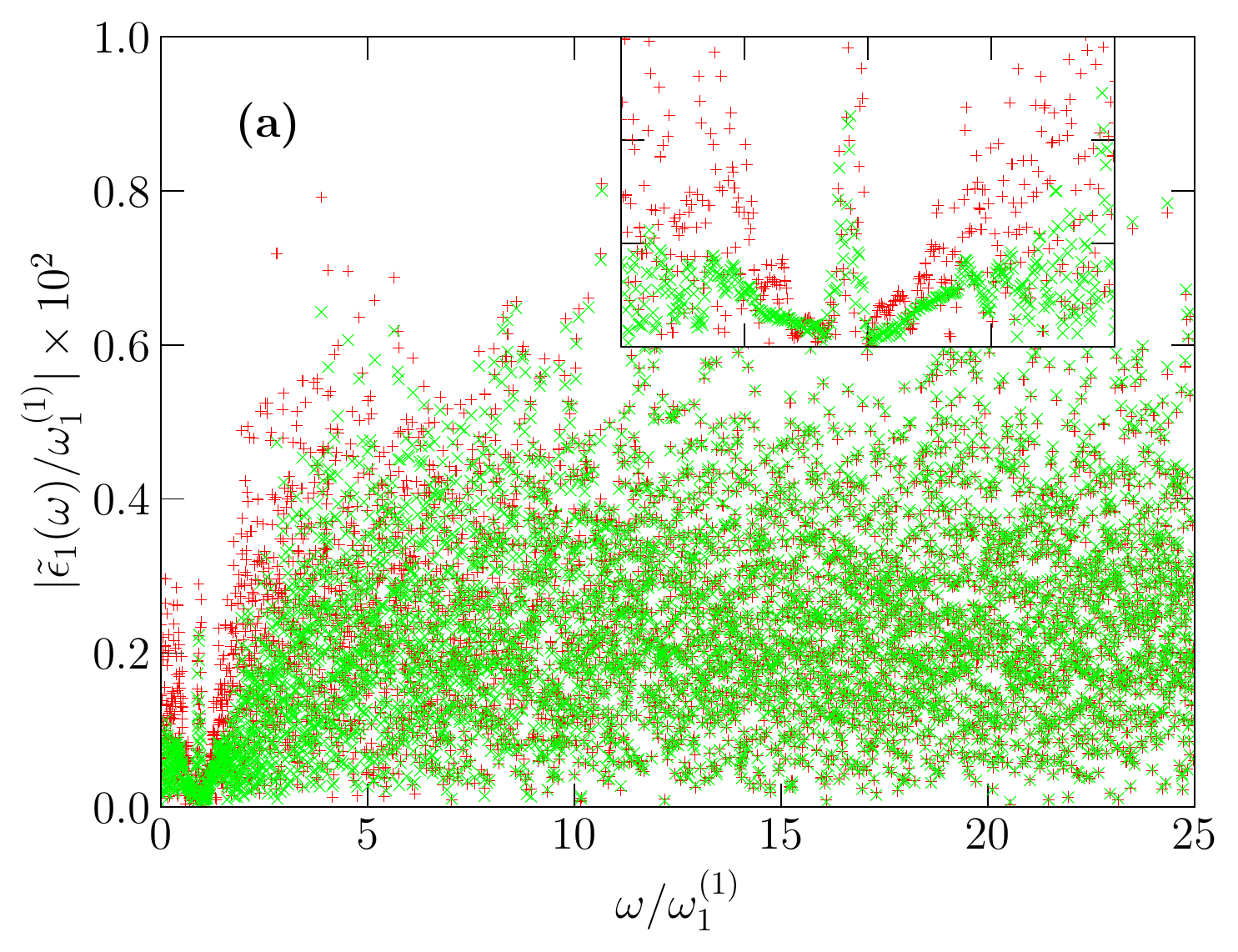}
\includegraphics[width=8.0cm]{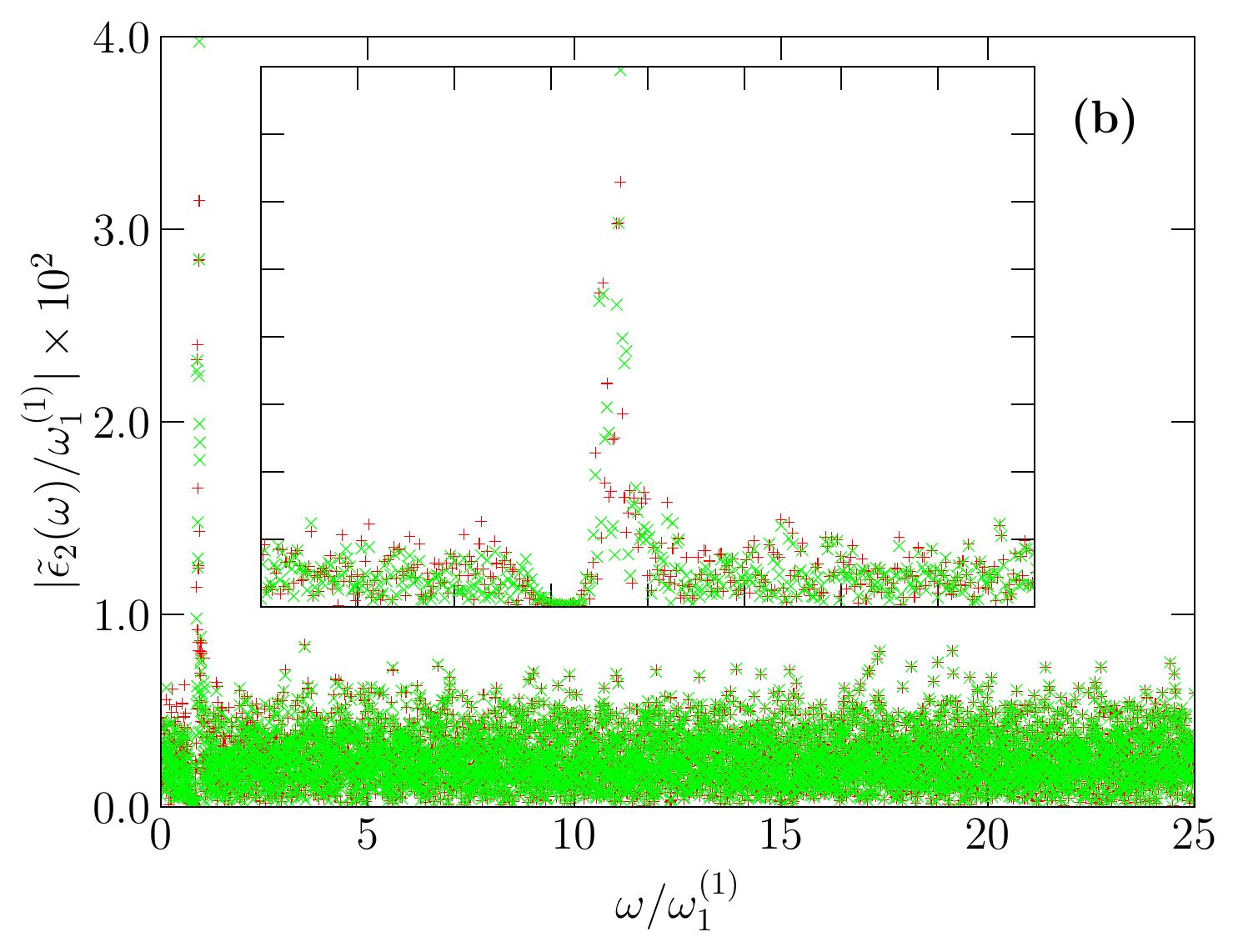}
\caption{Absolute values of the Fourier coefficients $\tilde{\epsilon}_k(\omega)$ of the optimized pulses $\epsilon_k(t)$ for the two qubits ($k=1,2$) as functions of frequency $\omega$ for the case shown in Fig.~\ref{Fig:OCTVaryingLossRates}(b). We plot only the positive frequencies, because the Fourier transform is symmetric with respect to the sign of $\omega$. The red + symbols are obtained with our standard $10^4$ optimization iterations, while the green $\times$ symbols are obtained after $5\times10^{4}$ iterations. The insets show magnified views of the low-frequency part of the spectrum: in both insets the x-axis range is [0,2], while the y-axis range is [0,0.3] in (a) and [0,4] in (b).}
\label{Fig:OptimizedPulses}
\end{figure}

In Fig.~\ref{Fig:OptimizedPulses} we plot the Fourier transforms of the driving fields that correspond to Fig.~\ref{Fig:OCTVaryingLossRates}(b). We do not plot the signals as functions of time, i.e.~$\epsilon_k(t)$, because they look like noise signals with no discernible features. For the pulse that produces Fig.~\ref{Fig:OCTZeroLoss}(b), both the time-domain signals and their Fourier transforms are almost featureless. The Fourier transforms of the initial, randomly generated pulses also look like white-noise signals. 

Figure \ref{Fig:OptimizedPulses} shows peaks (with some internal features) in the frequency range 0.9-1, i.e.~at the scale of the qubit Larmor frequencies. Such peaks are to be expected for a system manipulated by resonant driving of its various transitions. An interesting observation here is that the peak in $|\tilde{\epsilon}_2(\omega)|$ is about five times higher than the peak in $|\tilde{\epsilon}_1(\omega)|$. This result appears to be inconsistent with the driving conditions needed for the CR/SD gate, in which the driving amplitude applied to the control qubit is significantly larger than that applied to the target qubit. We shall return to this point in Sec.~\ref{Sec:CRSD} and show that this relation between the two signals has a possible logical explanation. If we move slightly away from the peaks, the Fourier coefficients of the control pulses are particularly small. Then, if we move farther away from the peaks, the Fourier coefficients become large again. In fact, their magnitude far away from the peaks is essentially the same as that in the initial guess pulse. This pattern indicates that the optimization algorithm is most effective in shaping the control pulse at relatively low frequencies, especially around the frequencies of the various transitions in the system. To demonstrate this point further, we continue the pulse optimization procedure for a total of $5\times 10^4$ iterations. Especially in Fig.~\ref{Fig:OptimizedPulses}(a), the low-frequency non-resonant components are significantly suppressed by the additional optimization, while the higher-frequency components are barely affected. This behavior is not too surprising, considering that the high-frequency components have a small effect on the long-time dynamics, such that they are given a low priority for refinement by the optimization algorithm. We expect that if we increase the number of iterations the algorithm will eventually suppress the high-frequency components of the signal and generate rather smooth control signals. However, achieving this goal using the GRAPE algorithm can take a prohibitively long computation time. In this context it is worth noting that the fidelity after $10^4$ iterations is 98.47\%, and it rises only slightly (to 98.73\%) after $5\times 10^4$ iterations, which shows that for the purpose of determining the speed limit, it is not necessary to obtain smooth pulses. For practical realizations of quantum gates in experiment, it is of course necessary to identify easily implementable pulses. Considering our discussion above about the minimal effect of high-frequency components, one can intuitively expect that taking the control signals in Fig.~\ref{Fig:OptimizedPulses} and filtering out the high-frequency components can quickly generate greatly optimized pulses. We filtered out low- and high-frequency components with varying cutoff frequency combinations. The fidelity generally remained high when we filtered out frequency components with $\omega\gtrsim 1.3 \ \omega_1^{(1)}$. The fidelity was more sensitive to filtering out low-frequency components. We do not show any of the resulting time-domain pulses here, because the pulses that gave high fidelities consistently looked noisy for our choice of system parameters.

\begin{figure}[h]
\includegraphics[width=8.0cm]{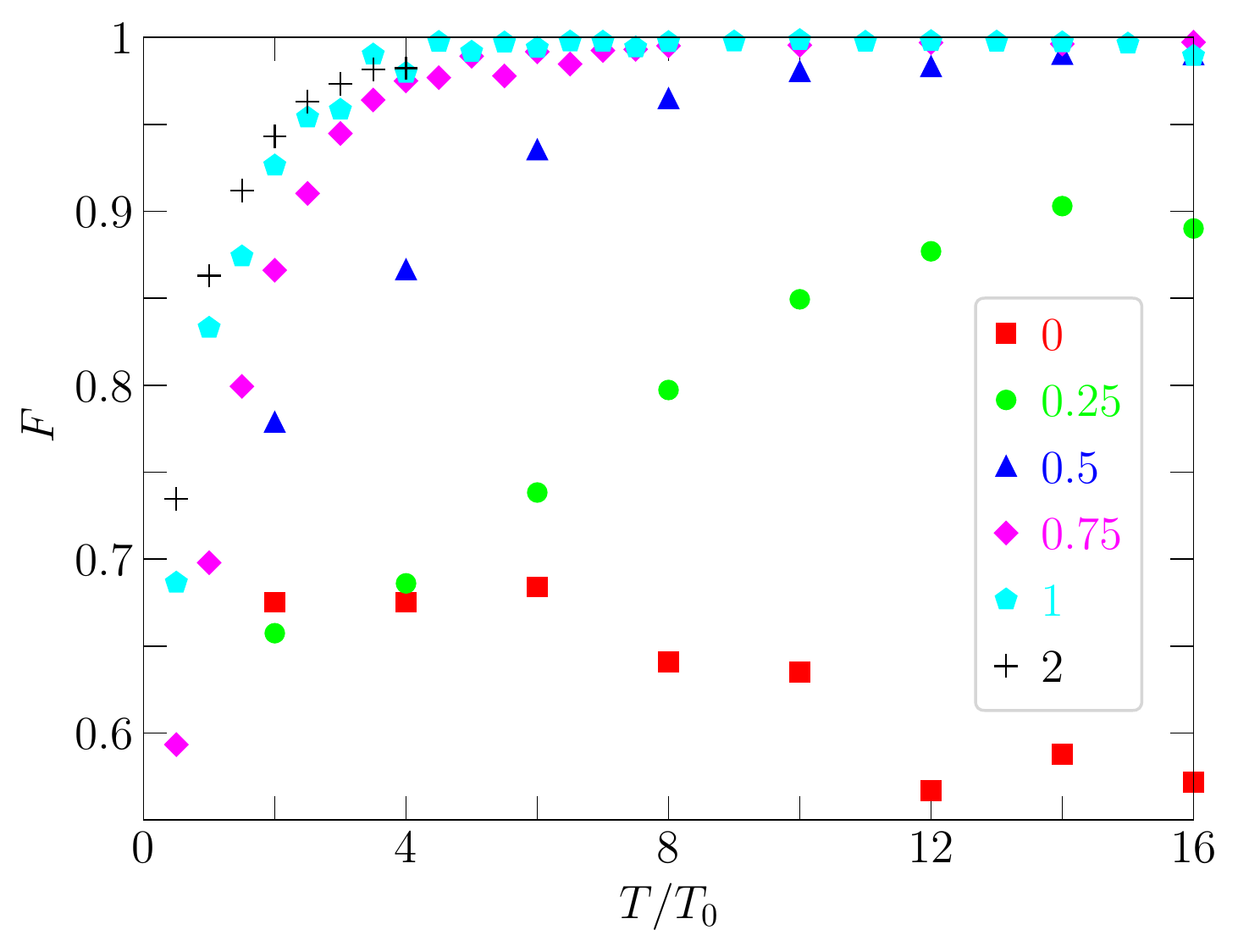}
\caption{Maximum fidelity as a function of pulse time,  as in Fig.~\ref{Fig:OCTZeroLoss}(a), but including the loss factor and varying the anharmonicity. In all cases, we keep four energy levels per qubit. The red squares, green circles, blue triangles, magenta diamonds, cyan pentagons and black + symbols correspond, respectively, to anharmonicity combinations $(\eta_3,\eta_4)=(-0.11,-0.19)\times\eta$ with $\eta=0$, $0.25$, $0.5$, $0.75$, $1$ and $2$. We use $(\gamma_3,\gamma_4)=(0,10^{-2})$ for all data sets. The same values of loss rate and anharmonicity are used for both qubits.}
\label{Fig:OCTVaryingAnharmonicity}
\end{figure}

Next we analyze the dependence of the speed limit on anharmonicity. In Fig.~\ref{Fig:OCTVaryingAnharmonicity} we plot the fidelity as a function of pulse time for different sets of anharmonicity values. For weak anharmonicities below $0.1$, the minimum gate time is roughly inversely proportional to the anharmonicity. This trend is indeed what one might intuitively expect, because weak anharmonicity makes it more difficult to address transitions among the lowest levels separately from transitions to higher levels. This trend is also consistent with past results on performing two-qubit gates with weakly anharmonic qubits \cite{Zhu,DeGroot2012}. In the extreme case of zero anharmonicity, it becomes practically impossible to achieve high-fidelity gates, because any driving that induces the $\ket{0}\leftrightarrow\ket{1}$ transition of a harmonic oscillator will also excite the system to higher levels where loss occurs. This dependence on the anharmonicity also contrasts with the near absence of any dependence on anharmonicity that we found in the calculations with no loss factor. The dependence on anharmonicity becomes weaker as the anharmonicity increases, which is also to be expected as stronger anharmonicities gradually make it increasingly easy to isolate the qubit space from the rest of the Hilbert space.

Before closing this section, we reconsider our results from a more general circuit design perspective. The existence of the additional quantum states opens new channels for the dynamics and enables the possibility of achieving faster quantum gates. The weak anharmonicity can complicate the targeted control of the different transitions in the energy level ladder. It is therefore desirable to have qubits that have additional energy levels in the realistically accessible range, e.g.~with transition frequencies on the scale of a few GHz for superconducting qubits, but are not weakly anharmonic. For this purpose, qubit designs such as the capacitively shunted flux qubit \cite{You,Steffen,Yan} or the fluxonium \cite{Manucharyan} can provide an advantage in terms of the achievable gate speed, because these qubit designs are more strongly anharmonic than the phase qubit and transmon designs. As a result, they can allow a more controlled utilization of any additional quantum states outside the qubit space, as demonstrated recently in experiments on qutrit control in a capacitively shunted flux qubit device \cite{Yurtalan,Kononenko}. As mentioned above, higher energy levels generally correspond to more delocalized states when considering the wave functions in terms of the circuit variables. It can therefore be expected that higher levels will generally lead to stronger coupling than the lowest energy levels. With these considerations in mind, the ability to utilize higher levels to speed up quantum operations can serve as an additional metric when assessing new qubit designs.

\section{Cross-resonance/selective-darkening gate}
\label{Sec:CRSD}

There are a few studies in the literature on the effect of higher levels on specific implementations of two-qubit gates \cite{Zhu,DeGroot2012,Lu,Ghosh,Kirchhoff,Tripathi,Magesan}. These previous studies have generally considered the effect of leakage to the higher levels and energy level shifts caused by the combination of driving and higher levels. In fact these energy level shifts can be highly non-negligible, as was shown in recent experiments on qutrit gates \cite{Yurtalan}. It can be expected that the deleterious effects of the higher levels worsen with increasing driving strength, which is intuitively associated with faster gates. We shall see below that, although this trend is obtained for unoptimized pulses, relatively simple optimization of the parameters can in some cases lead to a significant improvement in the gate fidelity with speeds not far below the speed limits.

For comparison with the OCT results presented in Sec.~\ref{Sec:Results}, we perform calculations similar to those reported in Ref.~\cite{DeGroot2012} to analyze the performance metrics, including the gate speed, of the CR/SD gate with the system parameters used in this work. The picture that one would expect from this kind of calculation is as follows: as we increase the driving strength, the gate speed increases while the fidelity decreases. The increase in gate speed follows from the increase in the relevant transition matrix element, which is proportional to the driving amplitude. The decrease in fidelity is expected because the driving protocol is designed with the assumption that there are only two energy levels per qubit, and there is no correction mechanism in the driving protocol to deal with higher levels. The leakage and energy level shifts caused by the higher energy levels then result in deviations from the desired gate dynamics, which leads to lower gate fidelities. In such a situation, where faster gates correlate with lower fidelities, one typically decides in advance what minimum fidelity is required or desired, and one chooses the gate speed that corresponds to this minimum fidelity.

The first set of calculations in this section proceed similarly to those of Ref.~\cite{DeGroot2012}. We perform simulations of the driven system dynamics, keeping either three or four energy levels for each qubit. In the CR/SD protocol, the system is driven at the frequency of the target qubit, which we take to be qubit 2. With a properly chosen combination of pulse amplitude and duration, a CNOT gate (or an equivalent two-qubit gate) is obtained. We use the SD implementation of the gate, i.e.~the two qubits are driven simultaneously such that the $\ket{00}\leftrightarrow\ket{01}$ transition is completely suppressed. We assume a pulse envelope shaped as the sine function $\sin(x)$ from $x=0$ to $x=\pi$. In other words, the driving fields are given by $\epsilon_1(t)=\epsilon_{\rm max} \sin(\pi t/T) \cos \left( \omega_1^{(2)} t \right)$ and $\epsilon_2(t)$ calculated accordingly, where $\epsilon_{\rm max}$ is the maximum value of the driving field amplitude on qubit 1 and is used to quantify the driving strength, and $T$ is the pulse duration. For each value of driving strength, we first estimate the corresponding pulse duration $T_e$ based on the matrix element for the CNOT gate transition. The matrix element is obtained by numerically diagonalizing the $9\times 9$ or $16\times 16$ Hamiltonian, depending on whether we keep 3 or 4 energy levels per qubit. We then simulate the driven dynamics for 200 values of the pulse duration ranging from zero to $6T_e$. For each one of these values for the pulse duration, we evaluate the fidelity of the implemented unitary operator with the ideal CNOT gate. For the fidelity calculation, we search the space of single-qubit unitary operators for operations that can be applied before and/or after the CNOT gate pulse to maximize the fidelity. With 200 pulse duration values and their corresponding fidelity values at hand, we inspect the fidelity values starting from zero pulse duration and moving up. The fidelity exhibits oscillatory behavior, characterized by a sequence of peaks and dips, as a function of pulse duration. If we reach a peak in the fidelity that is above 0.99, we identify the peak location as the gate time. In some cases, especially when some undesired resonance occurs, no high-fidelity peak is encountered in the range of pulse durations from zero to $6T_e$. In these cases, we take the pulse duration that corresponds to the highest fidelity and use it as the gate time.

\begin{figure}[h]
\includegraphics[width=8.0cm]{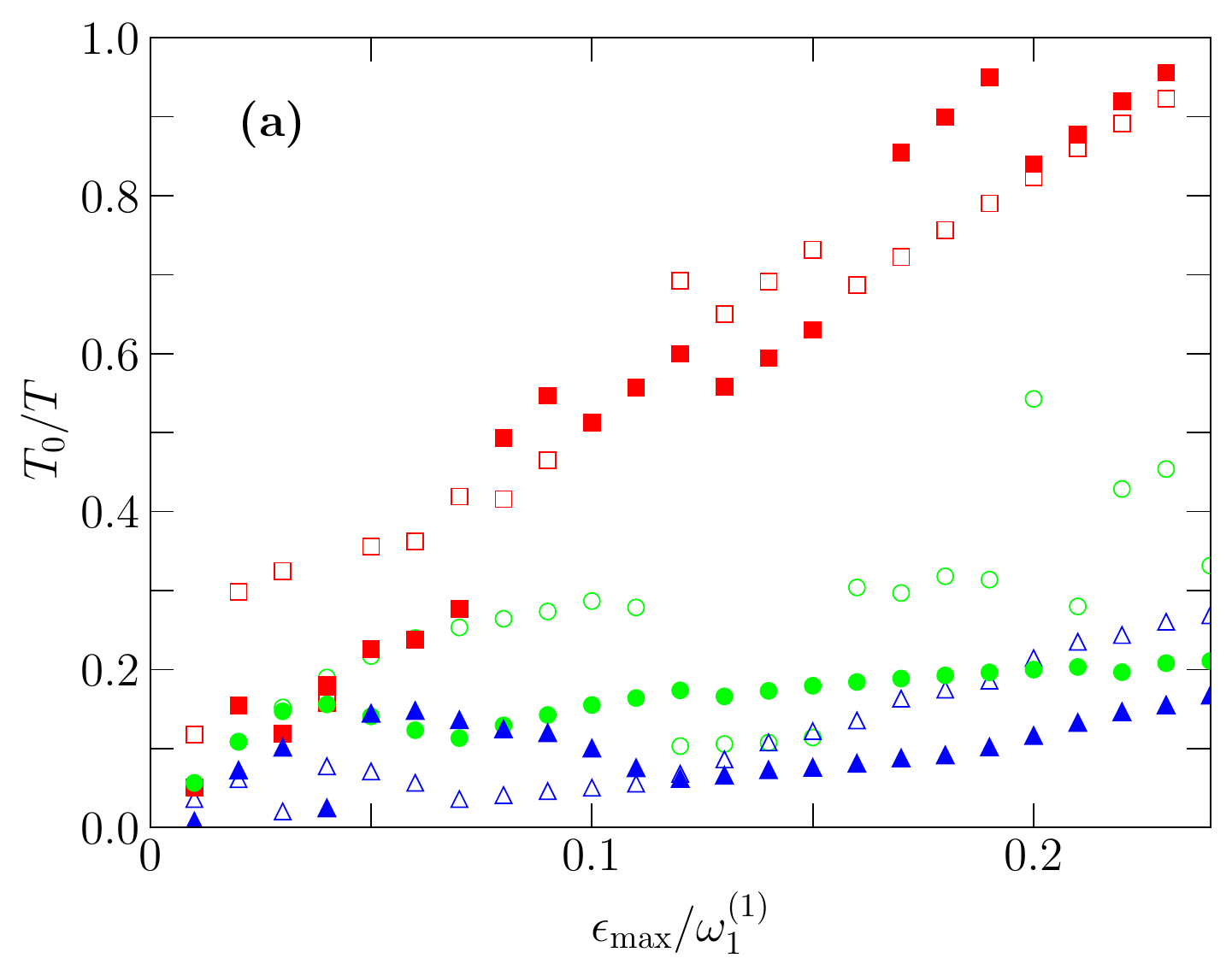}
\includegraphics[width=8.0cm]{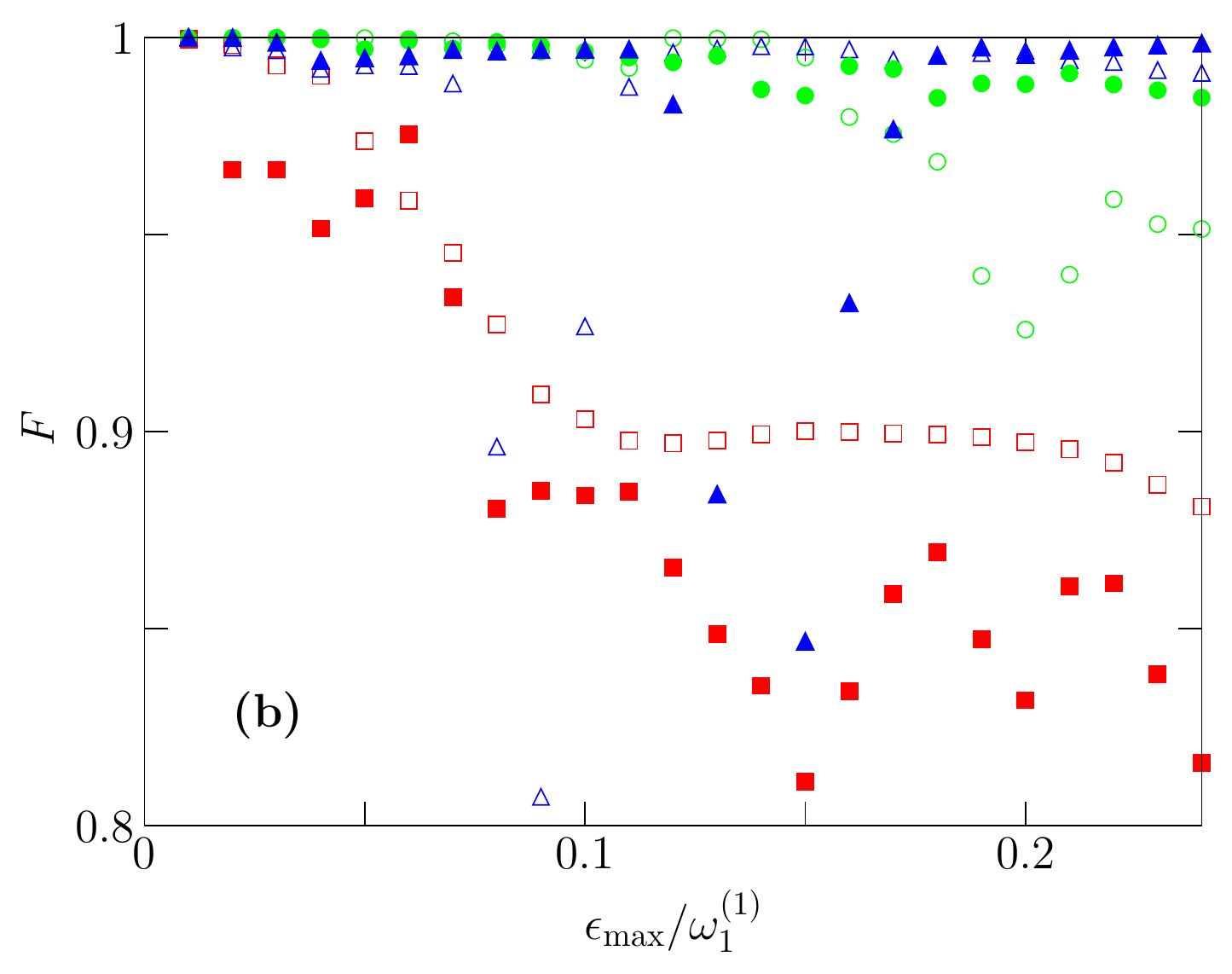}
\caption{Gate speed $T_0/T$ and fidelity $F$ as functions of driving strength $\epsilon_{\rm max}$ (measured relative to $\omega_1^{(1)}$) for the CR/SD gate. The red squares, green circles and blue triangles correspond, respectively, to $\eta_3=-0.11$, $-0.22$ and $-0.055$. The open symbols correspond to keeping three energy levels per qubit, while the closed symbols correspond to keeping four levels per qubit. The ratio $T_0/T$ is used as the gate speed quantifier because it is equal to 1 at the speed limit for simple qubits.}
\label{Fig:SelectiveDarkeningNoDecay}
\end{figure}

The results for the gate speed and the corresponding gate fidelity are shown in Fig.~\ref{Fig:SelectiveDarkeningNoDecay}. It might look surprising that in the data sets for $\eta_3=-0.055$ and $\eta_3=-0.22$ the fidelity remains close to 1 and does not decrease with increasing driving strength. The reason behind these high fidelities is that in this calculation we have allowed for numerically optimized single-qubit rotations to be applied before and after the CR/SD gate pulse, in addition to varying the pulse duration and choosing the value that gives the highest fidelity. This procedure amounts to performing a pulse optimization algorithm, with a much smaller number of variable parameters compared to GRAPE but with simple pulses that are guaranteed to be easily implementable. If we simply follow the basic theoretical formulae for the CR/SD pulse parameters when designing the CNOT gate pulse, e.g.~as in Refs.~\cite{Rigetti,DeGroot2010}, the fidelity would remain low throughout the plotted range. Even for weak driving, where leakage can be avoided, the energy level shifts caused by the higher levels prevent a straightforward implementation of the CR/SD gate in its basic form.

The results in the case $\eta_3=-0.11$ partly follow the simple picture described at the beginning of this section, namely an increase in gate speed accompanied by a decrease in fidelity. While the gate speed increases and approaches the simple-qubit speed limit, the fidelity drops significantly below 1, making these fast gates of little value for practical use in a quantum computing device. The fast deterioration of fidelity in this case can be understood by noting that the frequency $\omega_2^{(1)}-\omega_1^{(1)}=0.89$, which corresponds to a leakage transition that takes the system outside the computational space, is very close to the CR/SD driving frequency $\omega_1^{(2)}=0.9$. As a result, the standard CR/SD protocol fails badly. Somewhat surprisingly, the other two values of anharmonicity used in these calculations do not suffer a serious deterioration in the fidelity. This feature could be due to the slow rise and fall in the gate pulse amplitude, which can lead to an adiabatic population and depopulation of the higher levels over the course of the gate dynamics. There are, however, localized resonances where the fidelity exhibits dips at certain values of the driving strength.

Another unexpected feature is seen most clearly for the three-level simulations with $\eta_3=-0.11$. The gate speed does not vanish when the driving amplitude is reduced to zero, even though no driving-induced oscillations occur in this case. The cause of this undriven gate dynamics is that the higher levels cause qubit-state-dependent energy level shifts in the qubit space, and these shifts can induce entangling dynamics even in the absence of driving \cite{DeGroot2012}. As a result, one can obtain a two-qubit transformation that is equivalent to a CNOT gate in a finite amount of time even with no external driving. In order to establish a system for universal quantum computing, one must be able to suppress this always-on entangling dynamics when it is not needed. Otherwise it would be an especially serious issue for performing quantum information protocols using systems with strong coupling. Using OCT methods, it was shown in Ref.~\cite{Goerz} that one can controllably activate and suppress the entangling dynamics in these systems, such that universal quantum computing is possible once the pulses needed to implement the different operations are determined and properly applied.

We can now use Fig.~\ref{Fig:SelectiveDarkeningNoDecay} to infer information about the speed limit for the CR/SD gate. By looking for the largest value of $T_0/T$ that corresponds to a high fidelity, one can extract a minimum gate time for each set of system parameters. In spite of the differences between the different data sets in Fig.~\ref{Fig:SelectiveDarkeningNoDecay}, the maximum value of $T_0/T$ that is associated with a high fidelity is consistently around 0.15-0.25. The minimum gate time in these cases is therefore $\sim 4$-$6 T_0$, which is an order of magnitude longer than the minimum gate times obtained in Sec.~\ref{Sec:Results} in the zero-loss case ($\sim T_0/2$).

\begin{figure}[h]
\includegraphics[width=8.0cm]{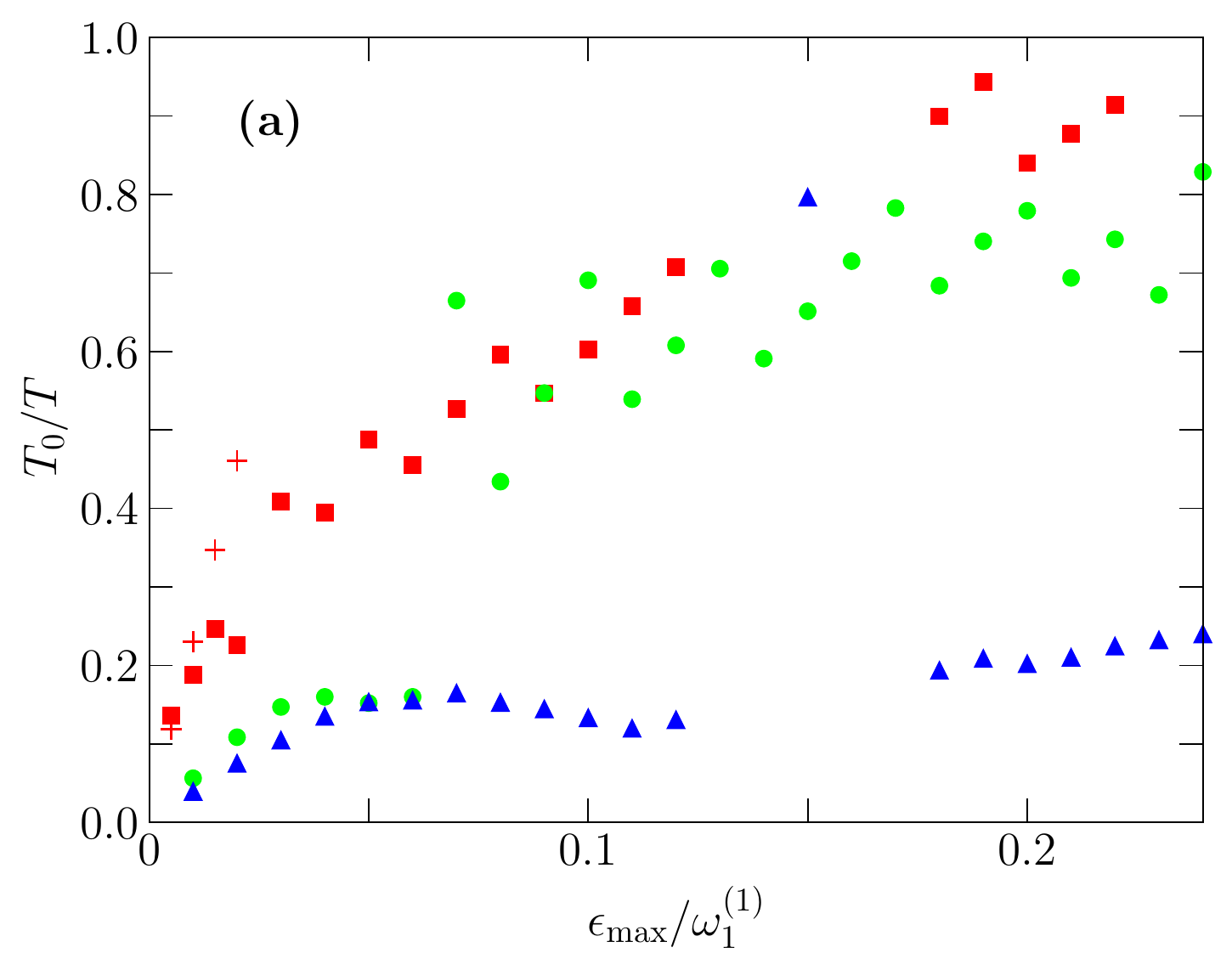}
\includegraphics[width=8.0cm]{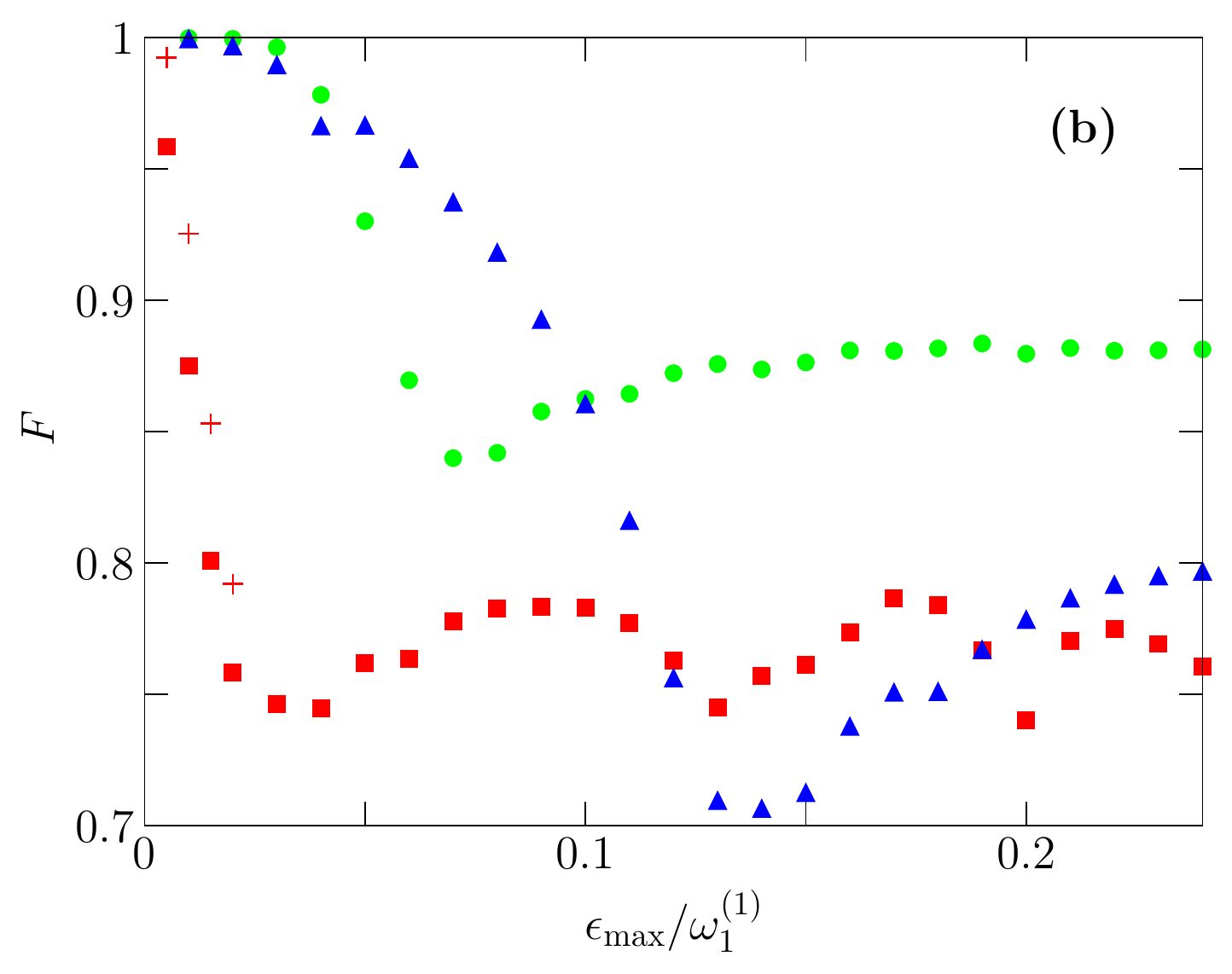}
\caption{Same as in Fig.~\ref{Fig:SelectiveDarkeningNoDecay}, but including loss from the higher levels. Here only four-level simulations are performed, with decay rates $(\gamma_2,\gamma_3)=(0,0.01)$. The + symbols plotted for low values of $\epsilon_{\rm max}$ are obtained by driving the qubits with an amplitude ratio designed to darken the $\ket{11}\leftrightarrow\ket{21}$ transition and hence suppress the leakage through that transition. To make the comparison between the squares and + symbols clear, there are more of these symbols than circles and triangles at low values of $\epsilon_{\rm max}$. Some data points seem to be missing in the gate speed plot. These points lie outside the plot range, i.e.~they have $T_0/T>1$, and they are ignored because they correspond to low fidelities.}
\label{Fig:SelectiveDarkeningWithDecay}
\end{figure}

Another feature that we can see in Fig.~\ref{Fig:SelectiveDarkeningNoDecay} is the substantial difference between the three- and four-level results, which indicates that the higher levels are significantly populated during the dynamics. We confirm this behavior in the population dynamics (not shown in the figure). As in Sec.~\ref{Sec:Results}, we must be cautious when interpreting simulation results that involve a significant population in the highest levels, which would indicate that even higher levels might need to be included in the simulation. To deal with complications that could arise in this case, we perform further simulations where we now include loss from the highest levels, such that any gate implementation that involves a significant population of these levels would result in a low fidelity and be disqualified in the search for optimal gate implementations. The results are shown in Fig.~\ref{Fig:SelectiveDarkeningWithDecay}. We can see clearly that the fidelity is high only when the driving is weak. In the case $\eta_3=-0.11$, even the weakest-driving data point in the figure has a low fidelity. Considering only the high-fidelity cases (e.g.~$F>0.99$), we obtain a maximum gate speed of about $T_0/T\sim$ 0.07-0.12, i.e.~a minimum gate time of $T\sim$ 8-15 $T_0$, for the three data sets shown in Fig.~\ref{Fig:SelectiveDarkeningWithDecay}.

It is worth considering the case $\eta_3=-0.11$ a little bit further. The reason for the great reduction in fidelity in this case is the leakage from the state $\ket{1}$ to the state $\ket{2}$ in qubit 1. As explained in Ref.~\cite{DeGroot2012}, the CR/SD gate can be performed using any combination of driving amplitudes on the two qubits. We therefore perform additional simulations where we set the amplitudes using the condition that the $\ket{11}\leftrightarrow\ket{21}$ transition is darkened to suppress the leakage through this transition. To achieve complete darkening of the $\ket{11}\leftrightarrow\ket{21}$ transition, the ratio between the driving amplitudes $\epsilon_2(t)/\epsilon_1(t)$ should be set to $-3.4$. This ratio is comparable to that obtained from the OCT calculations that produced Fig.~\ref{Fig:OptimizedPulses}, which could provide a logical interpretation for the amplitude ratio in the OCT results. With the ratio $-3.4$ used for the CR/SD driving amplitudes, we find that the fidelity deterioration with increasing driving amplitude is slowed down, such that we can achieve a gate time comparable to those obtained for the other anharmonicity values.

As expected, the results obtained in this section when avoiding higher level excitation give slower gates ($\sim 10T_0$ with parameter dependent variations) compared to those obtained using OCT ($\sim 4T_0$ with variations). However, the difference (a factor of 2-3) is not as large as one might have expected, taking into consideration the much larger amount of freedom in designing pulses in OCT. This result means that in general one can to some extent approach the speed limit using the CR/SD protocol with a relatively simple pulse optimization procedure. Further improvements with the simple gate implementation could be obtained by incorporating leakage-suppression techniques such as DRAG. One clear exception where we did not obtain high fidelity gates with simple pulses is when we encountered undesired resonances. The OCT results did not exhibit any resonance features as in Fig.~\ref{Fig:SelectiveDarkeningNoDecay} and possibly Fig.~\ref{Fig:SelectiveDarkeningWithDecay}, which is to be expected, because OCT methods naturally recognize pulses that lead to unintended leakage resonances and move the search away from such pulses. As a result, even in the case $\eta_3=-0.11$, OCT shows that we can achieve typical gate speeds with arbitrarily high fidelities when using optimized pulses.

\section{Conclusion}
\label{Sec:Conclusion}

In conclusion, we have investigated the application of OCT methods to the problem of implementing two-qubit gates with weakly anharmonic qubits, in which case the search for optimal pulses allows the dynamics to involve the irrelevant part of the Hilbert space at intermediate times but can also be programmed to discourage the population of specific states. We have found that, depending on whether higher energy levels can be controlled at high speeds and with low loss or not, the presence of these extra levels can either speed up or slow down the implementation of two-qubit gates. Whether the higher levels are useful channels for faster gates or harmful leakage channels depends in part on the magnitude of the anharmonicity. More specifically, a moderately strong anharmonicity  that is typical for some superconducting qubit designs could allow the controlled utilization of the higher levels, pointing to a potential advantage of such qubit designs for achieving fast quantum gates. By comparing the results of OCT with a relatively simple optimization procedure for the CR/SD gate, we found that the latter can give high-fidelity gates with speeds not far below the speed limit, indicating that one can approach the speed limit with relatively simple control pulses.

The methods and results presented in this work can help guide future studies aiming to find optimal approaches to implement various operations in systems of weakly anharmonic qubits. They can also be adapted and utilized for the optimization of qudit operations in weakly anharmonic qudits \cite{Gokhale}. From a broader perspective, the idea of utilizing states outside the computational space to speed up quantum operation is based on general physical principles. It is therefore not limited to superconducting systems, and it could be applied in other quantum computing platforms.

\section*{Acknowledgments}

We would like to thank Mikio Fujiwara, Christiane Koch and Masahiro Takeoka for useful discussions. This work was supported by MEXT Quantum Leap Flagship Program Grant Number JPMXS0120319794 and by Japan Science and Technology Agency Core Research for Evolutionary Science and Technology Grant Number JPMJCR1775. AL acknowledges support from the National Sciences and Engineering Council of Canada (NSERC).


\begin{thebibliography}{99}

\bibitem{Ladd} T. D. Ladd, F. Jelezko, R. Laflamme, Y. Nakamura, C. Monroe, J. L. O'Brien, Nature {\bf 464}, 45 (2010).

\bibitem{Buluta} I. Buluta, S. Ashhab, and F. Nori, Rep. Prog. Phys. {\bf 74}, 104401 (2011).

\bibitem{Kjaergaard} M. Kjaergaard, M. E. Schwartz, J. Braum\"uller, P. Krantz, J. I.-J. Wang, S. Gustavsson, and W. D. Oliver, Annu. Rev. Condens. Matter Phys. {\bf 11}, 369 (2020).

\bibitem{AshhabSpeedLimits} S. Ashhab, P. C. de Groot, and F. Nori, Phys. Rev. A {\bf 85}, 052327 (2012).

\bibitem{Martinis} J. M. Martinis, S. Nam, J. Aumentado, and C. Urbina, Phys. Rev. Lett. {\bf 89}, 117901 (2002).

\bibitem{Motzoi} F. Motzoi, J. M. Gambetta, P. Rebentrost, and F. K. Wilhelm, Phys. Rev. Lett. {\bf 103}, 110501 (2009).

\bibitem{Strauch} F. W. Strauch, P. R. Johnson, A. J. Dragt, C. J. Lobb, J. R. Anderson, and F. C. Wellstood, Phys. Rev. Lett. {\bf 91}, 167005 (2003).

\bibitem{Matsuo} S. Matsuo, S. Ashhab, T. Fujii, F. Nori, K. Nagai, and N. Hatakenaka, J. Phys. Soc. Jpn. {\bf 76}, 054802 (2007).

\bibitem{Neeley} M. Neeley, M. Ansmann, R. C. Bialczak, M. Hofheinz, E. Lucero, A. D. O'Connell, D. Sank, H. Wang, J. Wenner, A. N. Cleland, M. R. Geller, and J. M. Martinis, Science {\bf 325}, 722 (2009).

\bibitem{DiCarlo} L. DiCarlo, J. M. Chow, J. M. Gambetta, Lev S. Bishop, B. R. Johnson, D. I. Schuster, J. Majer, A. Blais, L. Frunzio, S. M. Girvin, and R. J. Schoelkopf, Nature {\bf 460}, 240 (2009).

\bibitem{Solenov2014} D. Solenov, S. E. Economou, and T. L. Reinecke, Phys. Rev. B {\bf 89}, 155404 (2014).

\bibitem{Solenov2016} D. Solenov, Quantum Inf. Comput. {\bf 16}, 954 (2016).

\bibitem{Nesterov} K. N. Nesterov, I. V. Pechenezhskiy, C. Wang, V. E. Manucharyan, and M. G. Vavilov, 	Phys. Rev. A {\bf 98}, 030301(R) (2018).

\bibitem{Werschnik} J. Werschnik and E. K. U. Gross, J. Phys. B: At. Mol. Opt. Phys. {\bf 40}, R175 (2007).

\bibitem{Spoerl} A. Sp\"orl, T. Schulte-Herbr\"uggen, S. J. Glaser, V. Bergholm, M. J. Storcz, J. Ferber, and F. K. Wilhelm, Phys. Rev. A {\bf 75}, 012302 (2007).

\bibitem{Mueller} M. M. M\"uller, D. M. Reich, M. Murphy, H. Yuan, J. Vala, K. B. Whaley, T. Calarco, and C. P. Koch, Phys. Rev. A {\bf 84}, 042315 (2011).

\bibitem{Reich} D. M. Reich, M. Ndong, and C. P. Koch, J. Chem. Phys. {\bf 136}, 104103 (2012).

\bibitem{Huang} S.-Y. Huang and H.-S. Goan, Phys. Rev. A {\bf 90}, 012318 (2014).

\bibitem{Watts} P. Watts, J. Vala, M. M. M\"uller, T. Calarco, K. B. Whaley, D. M. Reich, M. H. Goerz, and C. P. Koch, Phys. Rev. A {\bf 91}, 062306 (2015); M. H. Goerz, G. Gualdi, D. M. Reich, C. P. Koch, F. Motzoi, K. B. Whaley, J. Vala, M. M. Müller, S. Montangero, and T. Calarco
Phys. Rev. A {\bf 91}, 062307 (2015).

\bibitem{Heeres} R. W. Heeres, P. Reinhold, N. Ofek, L. Frunzio, L. Jiang, M. H. Devoret, and R. J. Schoelkopf, Nature Commun. {\bf 8}, 94 (2017).

\bibitem{Hu} L. Hu, Y. Ma, W. Cai, X. Mu, Y. Xu, W. Wang, Y. Wu, H. Wang, Y. P. Song, C.-L. Zou, S. M. Girvin, L-M. Duan, and L. Sun, Nature Phys. {\bf 15}, 503 (2019).

\bibitem{Wu} X. Wu, S. L. Tomarken, N. A. Petersson, L. A. Martinez, Y. J. Rosen, and J. L. DuBois, Phys. Rev. Lett. {\bf 125}, 170502 (2020).

\bibitem{Zong} Z. Zong, Z. Sun, Z. Dong, C. Run, L. Xiang, Z. Zhan, Q. Wang, Y. Fei, Y. Wu, W. Jin, C. Xiao, Z. Jia, P. Duan, J. Wu, Y. Yin, and G. Guo, Phys. Rev. Applied {\bf 15}, 064005 (2021).

\bibitem{Koch} J. Koch, T. M. Yu, J. Gambetta, A. A. Houck, D. I. Schuster, J. Majer, A. Blais, M. H. Devoret, S. M. Girvin, R. J. Schoelkopf, Phys. Rev. A {\bf 76}, 042319 (2007).

\bibitem{Chow} J. M. Chow, J. M. Gambetta, A. D. Corcoles, S. T. Merkel, J. A. Smolin, C. Rigetti, S. Poletto, G. A. Keefe, M. B. Rothwell, J. R. Rozen, M. B. Ketchen, and M. Steffen, Phys. Rev. Lett. {\bf 109}, 060501 (2012).

\bibitem{Barends} R. Barends, J. Kelly, A. Megrant, D. Sank, E. Jeffrey, Y. Chen, Y. Yin, B. Chiaro, J. Mutus, C. Neill, P. O’Malley, P. Roushan, J. Wenner, T. C. White, A. N. Cleland, and J. M. Martinis, Phys. Rev. Lett. {\bf 111}, 080502 (2013).

\bibitem{Rigetti} C. Rigetti and M. Devoret, Phys. Rev. B 81, 134507 (2010).

\bibitem{DeGroot2010} P. C. de Groot, J. Lisenfeld, R. N. Schouten, S. Ashhab, A. Lupascu, C. J. P. M. Harmans, and J. E. Mooij, Nature Phys. {\bf 6}, 763 (2010).

\bibitem{Khaneja} N. Khaneja, T. Reiss, C. Kehlet, T. S. Herbr\"uggen, S. J. Glaser, J. Magn. Reson. {\bf 172}, 296 (2005).

\bibitem{Brif} C. Brif, R. Chakrabarti, and H. Rabitz, New J. Phys. {\bf 12}, 075008 (2010).

\bibitem{Rebentrost} P. Rebentrost and F. K. Wilhelm, Phys. Rev. B {\bf 79}, 060507(R) (2009).

\bibitem{Molmer} K. M\o{}lmer and Y. Castin, Quantum Semiclass. Opt. {\bf 8}, 49 (1996).

\bibitem{Goerz} M. H. Goerz, F. Motzoi, K. B. Whaley, and C. P. Koch, npj Quantum Inf. {\bf 3}, 1 (2017).

\bibitem{Khani} B. Khani, J. M. Gambetta, F. Motzoi, and F. K. Wilhelm, Phys. Scr. {\bf T137}, 014021 (2009).

\bibitem{Zhu} D. Zhu, T. Jaako, Q. He, and P. Rabl, Phys. Rev. Applied {\bf 16}, 014024 (2021).

\bibitem{DeGroot2012} P. C. de Groot, S. Ashhab, A. Lupascu, L. DiCarlo, F. Nori, C. J. P. M. Harmans, J. E. Mooij, New J. Phys. {\bf 14}, 073038 (2012).

\bibitem{You} J. Q. You, X. Hu, S. Ashhab, and F. Nori, Phys. Rev. B {\bf 75}, 140515(R) (2007).

\bibitem{Steffen} M. Steffen, S. Kumar, D. P. DiVincenzo, J. R. Rozen, G. A. Keefe, M. B. Rothwell, and M. B. Ketchen, Phys. Rev. Lett. {\bf 105}, 100502 (2010).

\bibitem{Yan} F. Yan, S. Gustavsson, A. Kamal, J. Birenbaum, A. P. Sears, D. Hover, T. J. Gudmundsen, D. Rosenberg, G. Samach, S. Weber, J. L. Yoder, T. P. Orlando, J. Clarke, A. J. Kerman, and W. D. Oliver, Nature Commun. {\bf 7}, 12964 (2016).

\bibitem{Manucharyan} V. E. Manucharyan, J. Koch, L. Glazman, M. Devoret, Science {\bf 326}, 113 (2009).

\bibitem{Yurtalan} M.~A.~Yurtalan, J.~Shi, M.~Kononenko, A.~Lupascu, and S.~Ashhab, Phys.~Rev.~Lett.~{\bf 125}, 180504 (2020).

\bibitem{Kononenko} M.~Kononenko, M.~A.~Yurtalan, S.~Ren, J.~Shi, S.~Ashhab, and A.~Lupascu, Phys.~Rev.~Research {\bf 3}, L042007 (2021).

\bibitem{Lu} X. Y. L\"u, S. Ashhab, W. Cui, R. Wu, and F. Nori, New J. Phys. {\bf 14}, 073041 (2012).

\bibitem{Ghosh} J. Ghosh, A. Galiautdinov, Z. Zhou, A. N. Korotkov, J. M. Martinis, and M. R. Geller, Phys. Rev. A {\bf 87}, 022309 (2013).

\bibitem{Kirchhoff} S. Kirchhoff, T. Ke\ss ler, P. J. Liebermann, E. Ass\'emat, S. Machnes, F. Motzoi, and F. K. Wilhelm, Phys. Rev. A {\bf 97}, 042348 (2018).

\bibitem{Tripathi} V. Tripathi, M. Khezri, and A. N. Korotkov, Phys. Rev. A {\bf 100}, 012301 (2019).

\bibitem{Magesan} E. Magesan and J. M. Gambetta, Phys. Rev. A {\bf 101}, 052308 (2020).

\bibitem{Gokhale} P. Gokhale, A. Javadi-Abhari, N. Earnest, Y. Shi, F. T. Chong, arXiv:2004.11205.

\end{thebibliography}
\end{document}